\documentclass[aps,twocolumn,pre]{revtex4-2}

\usepackage{graphicx}
\usepackage{amsmath,mathtools,esint}
\usepackage{color,ulem,soul,braket}
\usepackage{booktabs}

\begin{document}

\title{Coherence in the Leak and Storage Kurtosis control Ergotropy in  Quantum Batteries}
\author{Bitap Raj Thakuria}
\author{Trishna Kalita}
\author{Manash Jyoti Sarmah}
\author{Himangshu Prabal Goswami}
\email{hpg@gauhati.ac.in}
\affiliation{QuAinT Research Group, Department of Chemistry, Gauhati University, Gopinath Bordoloi Nagar, Jalukbari, Guwahati 781014, Assam, India}
\date{\today}

\begin{abstract} 

 We introduce a cavity coupled finite-quantum system which can act as a quantum battery by harnessing noise induced coherences. 
We apply the methodology of full counting statistics to capture higher-order fluctuations of quanta exchange in the storage station. Together with the thermodynamic parameters, the fluctuations constitute a training platform for unsupervised as well as supervised learning models in predicting  ergotropy. 
We identify a minimal predictive feature set from the battery's operating parameters that can  classify the ergotropy into different regimes with great accuracy.
Our results show that the usual quantum and thermodynamic variables are inadequate for the purpose of identifying high ergotropy regimes in isolation. Rather, it is the kurtosis of quanta exchange in the storage  and the noise-induced coherence in the leakage mode that become the dominant quantities in controlling the magnitude of ergotropy.  

\end{abstract}

\maketitle







\section{Introduction}
Quantum batteries are intrinsically open quantum systems with efficient storage of energy and retrieval of work \cite{Campaioli2024RMP}. Unlike classical batteries, quantum batteries operate by keeping controlled population distributions and coherences in the energy eigenbasis of the system.  An essential figure of merit for quantum batteries is the ergotropy \cite{Allahverdyan2004,Ali2024ErgotropyCapacityPRA}, the maximum work extractable from a quantum state using unitary operations apart from the charging, storage, and discharging processes \cite{Allahverdyan2004,Ahmadi2024PRL_NonreciprocalQB,alicki2013,malavazi2025chargepreservingoperationsquantumbatteries}. Ergotropy not only relies on energy and population but also on  coherence\cite{Campaioli2024RMP,Shi2022CoherenceWorkPRL,Kamin2020EntanglementCoherenceCharging, Francica2020,Cakmak2020, Caravelli2021}. Its definition relies on the distinction between active and passive macrostates.  Energy in passive states cannot be accessed directly without further operations, while active states are directly useful for work extraction\cite{Perarnau2015,Cakmak2020}. In order to accurately estimate the ergotropy, a complete knowledge of the system's density matrix is required,\cite{Hadipour2024} and care should be taken while handling the steadystate coherences generated through interactions due to driving  during charging and storage\cite{Francica2020,OULARABI2025131003}. 
The charging station in traditional quantum battery are usually coherent drives\cite{khoudiri2025coherencedrivenquantumbatterycharging} or engineered reservoirs \cite{ahmadisuper}, while the storage space needs population and coherence stabilization against leakage\cite{PhysRevResearch.2.013095}, decoherence\cite{Sen2023NoisyQB} and backflow \cite{Ahmadi2024PRL_NonreciprocalQB}. Backflow of energy from storage to the power supply are suppressed through nonreciprocal interactions which minimize operational losses \cite{Wang2024PRA_COQB,Ahmadi2024PRL_NonreciprocalQB}.
Within the open quantum formalism, incoherent harmonic reservoirs are also natural choices as charging and discharging ports for quantum batteries since directional energy flow due to multiple incoherent baths is a known feature \cite{Tirone2025NoisyQBEfficiency,Song2025SelfDischargeQB,Khoudiri2025CoherenceDrivenQATM}. Such baths offer a physically motivated means of injecting and withdrawing energy while being in a nonequilibrium steadystate.\cite{2023JPhA...56A5302K} In contrast to idealized coherent drives, incoherent baths can be designed to be used at various temperatures or occupation numbers provide an experimentally friendly platform.\cite{Centrone2023}

 Stabilizing of coherences and populations  in the storage can be achieved by designing reversible energy exchange channels \cite{Wang2024PRA_COQB,QUTE2024_CavityScheme,rzgc-h78v,Hadipour2025BathModulationQB}. We propose that, accommodating  a cavity-mediated coherent channel,  through the connection of the storage subspace is a potential storage stabilizer.  A quantum cavity is capable of imposing a controllable, local but nonreciprocal interaction between the charging and storage stations for maintaining a preferred direction of energy flow into the storage subspace, block  backflow \cite{Utsumi2010}and increase extractable work. The hybrid combination of incoherent stations and stabilization via the cavity constitutes an alternative platform for efficient quantum energy storage. Nevertheless, incoherent charging operations and stochastic coupling with reservoirs produce energy and particle current fluctuations \cite{Hernandez2021} that affect the populations, coherences   and eventually, the ergotropy \cite{Esposito2009}. Ergotropy can be erased and the fidelity of charging as well as storage are prone to minimization \cite{Touchette2009}, if the system is not controlled properly.  The full counting statistics (FCS) formalism \cite{Paulino2024PRL_AdiabaticFCS} offers a useful tool to access these fluctuations quantified by the cumulants of energy exchange. Combining the FCS with the knowledge of active and passive states, we can explore the extent of the role of fluctuations on the battery energetics.\cite{Joshi2025PRL_FCS_Ions}
\begin{figure*}[t]
\centering
\includegraphics[width=\textwidth]{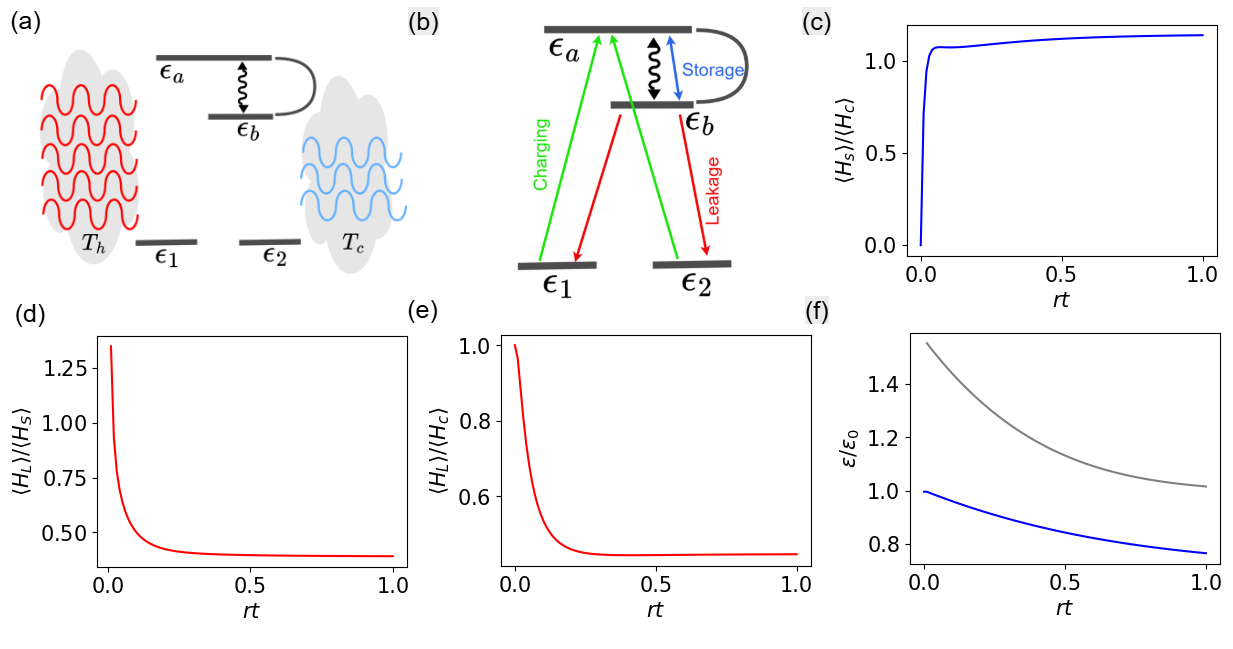}
\caption{
(a) Schematic representation of the cavity-mediated quantum battery model.
(b) Illustration of the charging, leakage, and storage pathways.
(c--e) Time evolution of the storage-to-charging, leakage-to-storage, and leakage-to-charging ratios for the parameters 
$\epsilon_1=\epsilon_2=0.1$, $\epsilon_b=0.4$, $\epsilon_a=1.5$, $r=1$, $g=1$, $p_h=0.61$, $p_c=0.97$, $T_c=5$, $T_\ell=1$, $T_h=6.36$, and $\tau=0.95$.
(f) Normalized ergotropy  \(\mathcal{E} / \mathcal{E}_{0}\) plotted as a function of the scaled time $rt$. 
The blue curve corresponds to the parameters used in panels (c--e), with the reference ergotropy  \(\mathcal{E}_{0}\) computed at $p_c=p_h=0$ and $\tau=0$. 
The grey curve uses the same parameters as (c--e) except $p_h=0.9$, $p_c=0.1$, $T_c=0.1$, and $T_h=2$, with  \(\mathcal{E}_{0}\) evaluated at $p_c=p_h=0$ and $\tau=0$.
All quantities shown are dimensionless.
}

\label{fig:battery}
\end{figure*}
The exploratory space of quantum batteries, or in general open quantum systems, is quite large, comprising of  temperatures, coupling strengths, energy level spacings, detunings, and cavity properties.  A comprehensive analytical or numerical investigation is  inconvenient. This problem has motivated the creation of data-driven methods, specifically machine learning (ML) algorithms, that are effective at finding optimal operation regimes. Through the generation of synthetic data sets over pertinent parameters, together with important observables like stored energy, charging power, leakage rates, steady-state currents, and low-order cumulants, ML models can identify ideal conditions, acquire mappings from system parameters to ergotropy, and reveal concealed correlations between fluctuations, coherences, and work extraction \cite{Rodriguez2023, Erdman2024, Hoang2024PRR_VQErgo, Sentz2025}. The current work is based on the combination of ML with FCS and cavity-assisted optimization to provide a rigorous method for optimizing quantum batteries  that are based on  realistic nonequilibrium conditions\cite{Du2025MeasQB_PRR, Rodriguez2024OQC_NJP}. The paper is organized as follows. In Sec. II, we introduce the battery model, the dynamics along with the thermodynamic ergotropy and the FCS. In sec. III we outline the integrated theoretical methodology that combine the FCS with ML. Sec IV is composed of results and discussion after which we conclude.

\section{Cavity Mediated Quantum Battery Dynamics}

\subsection{Quantum Battery Model}

We introduce a cavity mediated quantum battery comprising a
four-level system, a unimodal cavity and two harmonic reservoirs as shown in Fig. (\ref{fig:battery}a). 
This system can be interpreted as a quantum battery based on the following arguments. The states $|1\rangle$, $|2\rangle$  and $|a\rangle$  serve as the charging ports, receiving energy from the left side's harmonic reservoir. States $|a\rangle$ and $|b\rangle$ form the storage subspace, with energy  exchanged between them through a unimodal cavity. Transitions from $|b\rangle$ to $|1\rangle$, $|2\rangle$ represent leakage,  a loss mechanism to the other harmonic reservoir on the right side of the figure. The charging and leakage stations are reservoirs, assumed to be in an initial thermal state akin to a hot bath or cold bath at temperatures  $T_h$ and $T_c$ respectively. The  cavity-mediated coupling between $|a\rangle$ and $|b\rangle$ enhance the stability and control of energy storage. The working is shown as a separate diagram in Fig. (\ref{fig:battery}b).
The total Hamiltonian of the battery is given by:
\[
\hat{H}_{T} = \hat{H}_{0} + \hat{V}_{sb} + \hat{V}_{sc},
\]
where $\hat{H}_0$ represents the free Hamiltonian of the system and the harmonic stations, $\hat{V}_{sb}$ describes the stochastic interactions with harmonic stations.  $\hat{V}_{sc}$ accounts for coupling via a cavity mode with energy $\epsilon_\ell$. The free Hamiltonian is:
\[
\hat H_0 = \sum_{\nu=1,2,a,b} \epsilon_{\nu} |\nu \rangle\langle \nu | + \sum_{k\in h,c} \epsilon_k \hat{a}_k^{\dag} \hat{a}_k + \epsilon_\ell \hat{a}_{\ell}^{\dag} \hat{a}_{\ell},
\]
where $|\nu\rangle$ denote the system energy eigenstates, and $\hat{a}_k, \hat{a}_\ell$ are the annihilation operators for the harmonic and cavity modes respectively. The  interaction Hamiltonian is:
\[
\hat{V}_{sb} = \sum_{x \in \{h,c\}} \sum_{i=1,2} \sum_{k = a,b} \Gamma_{ix} \hat{a}_x |k\rangle\langle i| + \text{h.c.},
\]
which mediates energy exchange between the states and the stations, enabling transitions from charging states $|1\rangle$, $|2\rangle$ to intermediate levels $|a\rangle$, $|b\rangle$. The  interaction Hamiltonian is:
\[
\hat{V}_{sc} = g \hat{a}_{\ell}^\dag |b\rangle\langle a| + \text{h.c.},
\]
which enables reversible and controlled transitions between the storage states $|a\rangle$ and $|b\rangle$ via the cavity mode. 
When compared to a new oscillator based quantum battery recently put forward \cite{Ahmadi2024PRL_NonreciprocalQB}, we have some similarities and some differences. Both the nonreciprocal harmonic-oscillator battery \cite{Ahmadi2024PRL_NonreciprocalQB} and the currently proposed cavity-mediated  quantum battery share the same common platform i.e a driven open quantum system with controlled storage and directed energy flow. In the oscillator model, the charger and battery are modeled as two coupled bosonic modes with coherent interaction strength. Energy is fed into the charger through a classical drive and channeled into the battery mode through this coherent coupling, with dissipation and drive asymmetry making the process effectively nonreciprocal. Energy prefers to flow from charger to battery but not vice versa. The four-level model described here, two bosonic stations are used as stochastic chargers to populate  levels $\ket{1}$ and $\ket{2}$, with one cavity mode mediating coherent coupling between storage states $\ket{a}$ and $\ket{b}$. The hybrid design combines reservoir-driven excitation (the analog of the classical drive in the oscillator case) with cavity-assisted reversible energy exchange and lead to a natural generalization of the oscillator battery \cite{Ahmadi2024PRL_NonreciprocalQB,QUTE2024_CavityScheme}. 
In this regard, the cavity-assisted model embraces both the coherent control feature of quantum-optical batteries and the statistical energy fluxes of key importance in quantum thermodynamics, providing a more unified and physically complete  description of quantum energy storage. Although the coherent Hamiltonian of the cavity-assisted four-level quantum battery,$
\hat{V}_{sc} = g\, \hat{a}_\ell^\dag \ket{b}\bra{a} + \text{h.c.}$ 
is formally reciprocal, coupling the storage states $\ket{a}$ and $\ket{b}$ symmetrically via the cavity mode, the system exhibits effective nonreciprocity in its overall energy dynamics. This arises from the asymmetric stochastic coupling to the thermal reservoirs, which selectively populate the lower charging states $\ket{1}$ and $\ket{2}$ and mediate transitions to the storage manifold through $\hat{V}_{sb}$. The resulting architecture establishes a thermodynamic bias: energy flows preferentially from the left to right, through the storage states and can subsequently be extracted via controlled discharge transitions, when reverse flow back into the reservoirs is strongly suppressed\cite{Ahmadi2024PRL_NonreciprocalQB,Santos2021SelfDischargePRE}. Consequently, even though the coherent interactions themselves are symmetric, the combination of dissipative and thermodynamic asymmetries enforce a functional directionality in energy transfer, effectively rendering the battery nonreciprocal. This nonreciprocal energy flow ensures stable and unidirectional storage. The formalism of full counting statistics (FCS), previously developed to address the quanta fluctuations during energy exchange \cite{PhysRevA.86.043843,PhysRevA.110.052214}, albeit in a different context (a heat engine), can be used to study the stochastic dynamics for the battery  too. The dynamics are well captured through a known population-coherence coupled quantum master equation. Building on the already established foundations, the reduced density matrix in Liouville space can be represented as a state vector composed of four population terms and a quantum coherence component:
\[
|\rho\rangle = \{\rho_{11},\, \rho_{22},\, \rho_{bb},\, \rho_{aa},\, \Re(\rho_{12})\}, \quad i \in \{1,2,a,b\}.
\]
Here, $\Re(\rho_{12})$ denotes the real part of the coherence between the lower two states $|1\rangle$ and $|2\rangle$. This coherence arises from reservoir-induced quantum interference effects, specifically due to the asymmetric coupling of the system's many-body states with the harmonic reservoirs. We are primarily interested in the storage mode and to quantify the statistics of quanta exchange  events, we introduce a twisted evolution operator \cite{Esposito2009} that governs the statistical dynamics at the storage station of the system's reduced density vector. This is captured by the equation 
$|\dot{\rho}(\lambda, t)\rangle = \breve{\mathcal{L}}(\lambda)\, |\rho(\lambda, t)\rangle,$
where $\lambda$ is the counting field that tracks the net number of quanta exchanged between the states $\ket a$, $\ket b$ and the unimodal cavity. The twisted superoperator $\breve{\mathcal{L}}(\lambda)$ serves as the moment-generating operator for the quanta exchange statistics, and reduces to the standard evolution superoperator in the limit $\lambda \to 0$. Quantitatively \cite{PhysRevA.86.043843,PhysRevA.110.052214}, $\breve{\mathcal{L}} (\lambda)=$ 
\begin{align}
\label{Louv-eq}
\begingroup
\setlength{\arraycolsep}{4pt}
\renewcommand{\arraystretch}{1.2}
\resizebox{\columnwidth}{!}{\ensuremath{%
\left(
\begin{array}{ccccc}
-\displaystyle\sum_{x}\Gamma_{1x} n_x & 0 & \Gamma_{1h}\tilde{n}_h & \Gamma_{1c}\tilde{n}_c & -2\Gamma_{12} \\
0 & -\displaystyle\sum_{x}\Gamma_{2x} n_x & \Gamma_{2h}\tilde{n}_h & \Gamma_{2c}\tilde{n}_c & -2\Gamma_{12} \\
\Gamma_{1c} n_c & \Gamma_{2c} n_c & -\Gamma_h \tilde{n}_h - g^2 \tilde{n}_\ell & g^2 n_\ell \mathrm{e}^{-\lambda} & \Gamma_{12c} n_c \\
\Gamma_{1h} n_h & \Gamma_{2h} n_h & g^2 \tilde{n}_\ell \mathrm{e}^{\lambda} & -g^2 n_\ell - \Gamma_c \tilde{n}_c & 2\Gamma_{12h} n_h \\
-\Gamma_{12} & -\Gamma_{12} & \Gamma_{12h}\tilde{n}_h & 2\Gamma_{12c}\tilde{n}_c & -\bar{g} - \tau
\end{array}
\right)
}}%
\endgroup
\end{align}
where, $\Gamma_x= \Gamma_{1x}+\Gamma_{2x}, x =h,c. \ \Gamma_{ix}=\pi\Omega|g_{ix}^{}|^2/2, i =1,2$ represent rates of transition between the states and are proportional to the modulus square of the transition dipole between the states 
$|1\rangle (|2\rangle)$ and $|a\rangle$ or $|b\rangle$, $\bar g = n_h(\Gamma_{1h}+\Gamma_{2h})/2+
 n_c(\Gamma_{1c}+\Gamma_{2c})/2$ and $\Gamma_{12}=(\Gamma_{12c}n_c+\Gamma_{12h}n_h)/2$ with $\tau$ being a dimensionless pure dephasing parameter \cite{scully2011quantum,PhysRevA.86.043843,PhysRevA.110.052214}. $n_{c(h)}=1/\{\exp(\beta_{c(h)}(E_{b(a)}-E_1))-1\}$ are the Bose-Einstein distributions with $\tilde n_x =1+n_x$.
The mixed term $\Gamma_{12x}$ quantifies the contribution of coherence arising from the simultaneous interaction of both lower states $|1\rangle$ and $|2\rangle$ with a common harmonic reservoir $x \in \{h, c\}$. This coherence is encoded in the real part of the off-diagonal density matrix element, $\Re(\rho_{12})$, and stems from quantum interference between different transition pathways mediated by the harmonic stations. Mathematically, the mixed coupling term is defined as $
\Gamma_{12x} = \frac{\pi \Omega}{2} |\Gamma_{1x} \Gamma_{2x}|^2 
= \sqrt{\Gamma_{1x} \Gamma_{2x} |\cos \theta_x|}$,
where $\Gamma_{ix}$ represents the individual coupling strengths between state $|i\rangle$ ($i=1,2$) and the reservoir $x$, and $\theta_x$ is the angle that characterizes the relative orientation between the dipole transition moments associated with the transitions $|1\rangle \rightarrow |a\rangle$ and $|2\rangle \rightarrow |a\rangle$ (or similarly for transitions to $|b\rangle$). These dipole transitions interfere, and the strength of this interference depends on the geometric alignment of the dipoles,  constructive when parallel ($\theta_x = 0$), and destructive when orthogonal ($\theta_x = \pi/2$). Following Refs.~\cite{scully2011quantum,PhysRevA.88.013842}, and assuming symmetric system-bath coupling, i.e., $\Gamma_{1x} = \Gamma_{2x} = r$, the term simplifies to $
\Gamma_{12x} = r p_x,
$
where $p_x = \sqrt{|\cos \theta_x|}$ is a dimensionless parameter that quantifies the coherence-inducing capacity of the bath $x$. These are typically referred to as noise-induced coherences \cite{scully2011quantum}. The values of $p_x$ range from 0 (no coherence, orthogonal dipoles) to 1 (maximum coherence, parallel dipoles). Thus, $p_h$ and $p_c$ serve as tunable parameters for the hot and cold bath-induced coherences, respectively \cite{scully2011quantum,PhysRevA.88.013842, PhysRevResearch.4.L032034,PhysRevA.86.043843,svidzinsky2012enhancing,PhysRevA.107.052217}. Importantly, while $\Gamma_{12x}$ controls the degree of coherence, the impact of this coherence on physical observables,  such as photon flux or power output,  is nontrivial and generally nonlinear. For instance, within a perturbative treatment, both the photon flux into the cavity and the output power of the engine have been shown to depend nonlinearly on $p_h$ and $p_c$ \cite{scully2011quantum,PhysRevA.88.013842}.  Estimation  of these requires comprehensive knowledge of the  microscopic details, including the structure and spectral properties of the thermal reservoirs. In most experimental setups, reservoirs are realized via externally applied laser fields \cite{PhysRevLett.119.050602}, and the angle $\theta_x$ is determined by the relative orientation of the laser polarizations. To account for pure dephasing of the coherence $\rho_{12}$, a dimensionless parameter $\tau$ is added phenomenologically to the last element in Eq. (\ref{Louv-eq}). 

The term noise-induced coherence is used because the coherence does not come from a controlled or coherent driving field, but from the random, fluctuating noise of the charging and discharging reservoirs itself. In this context, noise refers to irregular electromagnetic fluctuations during energy exchange between system and reservoir due to frequency components capable of overlapping with more than one transition. Where two transitions share parts of the same spectral range, the same  noise can drive both transitions at once. This overlap makes their effects correlated rather than independent, allowing interference between the transition pathways. The result is that the random noise of the reservoirs give rise  a kind of order emerging from disorder. This can be visualized physically as the baths pushing the system along slightly tilted, non-vertical paths between energy levels. In other words, the reservoirs, which typically wash out coherence, itself generates it by means of its randomly fluctuating fields. The word noise is related to the fact  that this type of coherence is not engineered from the outside or driven by a laser, but stems from the randomness  through shared spectral components and interference among transitions. The two terms, $p_h$ and $p_c$ are noise-induced coherences with the former being the charging station coherence and $p_c$ being the leakage mode coherence, or the coherence in the leak.

\subsection{Battery Performance Indicators}
To quantitatively demonstrate the working feasibility of the proposed quantum battery, we evaluate the expectation values of thermodynamically relevant observables, the charging energy, $\langle \hat H_c\rangle$, the  storage energy, $\langle \hat H_s\rangle$ and the expected leakage, $\langle \hat H_L\rangle$. The operators are defined as
\begin{align}
    \hat H_c &= \sum_{\nu=1,2,a} \epsilon_{\nu} |\nu \rangle\langle \nu | \\
     \hat H_s &= \sum_{\nu=a,b} \epsilon_{\nu} |\nu \rangle\langle \nu |\\
      \hat H_L &= \sum_{\nu=1,2,b} \epsilon_{\nu} |\nu \rangle\langle \nu |
\end{align}
The ratio of the expectation values of the above operators are shown in Fig. (\ref{fig:battery}c,d,e). Steady-state populations $(\rho_{aa}, \rho_{bb}, \rho_{11}, \rho_{22})$ and coherence $\rho_{12}$ are derived by solving $\breve{\cal L}\rho=\dot\rho$ with the expectation values evaluated over the resultant density matrix. Fig. (\ref{fig:battery}c) highlights that storage in the steadystate can exceed the charging. Fig. (\ref{fig:battery}d, e) shows how the leakage with respect to storage and charging reduces as the system approaches steadystate. Note that, the extent of increase or decrease of these expectation values depends on what parameters are chosen. We added these figures as a proof of concept, and these establish  that the proposed model is capable of acting as a battery.

We now focus on the heart of the work, the ergotropy $\mathcal{E}$ which quantifies the maximum work that can be extracted from a
quantum state by unitary operations and is thus a key figure of merit for
quantum batteries. 
The system Hamiltonian is $\hat H=\mathrm{diag}(\epsilon_1,\epsilon_2,\varepsilon_b,\varepsilon_a)$ and define the ergotropy as,
\begin{equation}
{\cal E}=\mathrm{Tr}(\hat H\hat \rho) - \mathrm{Tr}(\tilde{\varrho} \hat H).
\label{eq:ergotropy}
\end{equation}
 $\tilde\varrho$ is called the passive state obtained by rearranging the basis of the density matrix $\hat\rho$ such that the difference of Eq. (\ref{eq:ergotropy}) becomes maximum. To achieve this, the energy basis is always written from low energy to high energy. The density matrix in such a basis is 
\begin{equation}
\hat\rho=\begin{pmatrix}
\rho_{11} & \rho_{12} & 0 & 0\\
\rho_{21} & \rho_{22} & 0 & 0\\
0 & 0 & \rho_{bb} & 0\\
0 & 0 & 0 & \rho_{aa}
\end{pmatrix}.
\label{eq:rhos}
\end{equation}
In the proposed battery, we obtain, $\rho_{11}=\rho_{22}$ and $\rho_{12}=\rho_{21}$, from which the passive density matrix needs to be identified. The exact procedure to calculate the passive state for such population-coherence coupled dynamics has been documented for this model\cite{sarmah2024noise} by following the general procedure \cite{Francica2020}. We construct the passive state  through a unitary transformation that diagonalizes
the density matrix \( \hat{\rho}(t) \) and arranges the resultant eigenvalues
in descending order\cite{Francica2020,sarmah2024noise}. 
The top-left block of Eq.~\ref{eq:rhos} forms a symmetric Toeplitz matrix whose
eigenvalues are well known, \( \rho_{\pm}(t) = \rho_{11}(t) \pm \rho_{12}(t) \).
The passive state \( \tilde{\varrho}(t) \) can be uniquely determined by
arranging the diagonalised density matrix components
\( \{ \rho_{-}(t), \rho_{+}(t), \rho_{bb}(t), \rho_{aa}(t) \} \)
in descending order for each variation in the Hamiltonian parameters. 
In the absence of coherence ($p_c=p_h=0$ and $\tau=0$), Eq.~\eqref{eq:ergotropy} reduces to the baseline ergotropy, or the classical value
$\mathcal{E}_0$, defined analogously from the diagonal state.  The term, $\mathcal{E}_0$ has no contribution from the noise-induced coherence. The ratio $\mathcal{E}$/$\mathcal{E}_0$  has been extensively studied 
to capture the enhancement provided by quantum coherences\cite{sarmah2024noise}. In Fig. (\ref{fig:battery}f), we show how the ratio behaves as a function of the dimensioneless time ($rt$) for two different parameter regimes keeping the coherence values fixed. As can be seen from the two curves (one being greater than unity and the other being less than unity) in Fig. (\ref{fig:battery}f),  coherences can amplify as well as diminish the ergotropy. It motivates our further analysis which focuses on identifying an optimal set of battery parameters or variables that  lead to high values of ergotropy, or coherence-enhanced ergotropy.

\subsection{Thermodynamic and Statistical Quantities}
\begin{table}
\caption{Sampling intervals for the quantum battery parameters. Eight random values are generated for each parameter, with the energy-level ordering constraint $\epsilon < \varepsilon_b < \varepsilon_a$ enforced through random gap sampling.}
\label{tab:MCintervals}
\begin{ruledtabular}
\begin{tabular}{lll}
Symbol & Range & Physical role \\
\colrule
$T_c$        & $0.1\dots7$      & Leak mode temperature \\
$T_h$        & $0.1\dots7$      & Charging temperature \\
$T_{\ell}$   & $0.1\dots7$      & Storage mode temperature \\
$\epsilon$   & $0.01\dots2$     & Ground-state energy \\
$\epsilon_b$ & $\epsilon + \delta_1$, $\ \delta_1 \sim (0.01,2)$  & First excited energy \\
$\epsilon_a$ & $\epsilon_b + \delta_2$, $\ \delta_2 \sim (0.01,2)$ & Second excited energy \\
$p_c,p_h$    & $0.1\dots1$      & Degree of Coherence \\
$\tau$       & $0.01\dots2$     & Dephasing \\
\end{tabular}
\end{ruledtabular}
\end{table}


Beyond the ergotropy,  fluctuations of the energy exchange also carry
information about the system's behaviour.  We are primarily interested in capturing the fluctuations of quanta exchange in the storage mode, i.e between levels $\ket a$, $\ket b$ and the unimodal cavity.  In the
long-time limit or steadystate, the fluctuations get naturally quantified through the cumulant generating function  \cite{Esposito2009}, the dominant and real eigenvalue $S(\lambda)$ of the superoperator $\mathcal{L}(\lambda)$. 
The $i$th cumulant $j^{(i)}$ is given by the $i$th
derivative of $S(\lambda)$ evaluated at $\lambda=0$.
\begin{equation}
\label{eq-cum}
    j^{(i)} = \left. \frac{\partial^i S(\lambda)}{\partial \lambda^i} \right|_{\lambda=0},
    \quad i=1,2,3,4.
\end{equation}
$ j^{(1)}$ corresponds to the mean exchange, $ j^{(2)}$, is the variance, $ j^{(3)}$ is the skewness and $ j^{(4)}$ is the kurtosis of quanta exchange at the storage station. We use the coherence-unaffected values of the cumulants as scaling factors to define a dimensionless cumulant
\begin{equation}
    C^{(i)} = \frac{j^{(i)}}{j^{(i)}_0},
\end{equation}
where $j^{(i)}_0$ denotes the cumulant in the absence of coherences. 
Because an analytic
expression for $S(\lambda)$ is unavailable, we calculate these derivatives
numerically by discretising $\lambda$ and applying high   order centred finite
difference stencils for accuracy. 
Here $C^{(1)}, C^{(2)}, C^{(3)}, C^{(4)}$ correspond to the dimensionless mean, variance, skewness, and kurtosis, respectively. One can go beyond the fourth order, but we have previously established that the first four are sufficient for analysis\cite{SARMAH2023129135}.
The other popular and standard thermodynamic quantities, the useful work output $W$, thermodynamic affinity $F$,  charging energy $Q_h$, leak mode energy, $Q_c$, and system's most probable efficiency $\eta$ \cite{sarmah2024noise} are obtained from already known standard thermodynamic relations, $ Q_h = e_a - e,
Q_c = e_b - e,
W = (e_a - e) - k_B T_c \ln F, 
F = \tilde{n}_c \, n_h \, \tilde{n}_\ell/(n_c \, \tilde{n}_h \, n_\ell) $ and $
\eta = W/Q_h$ with $\epsilon_1=\epsilon_2=e$, derived and established in earlier works \cite{PhysRevA.88.013842,sarmah2024noise}.
\begin{figure*}
 \centering
        \includegraphics[width=1\textwidth]{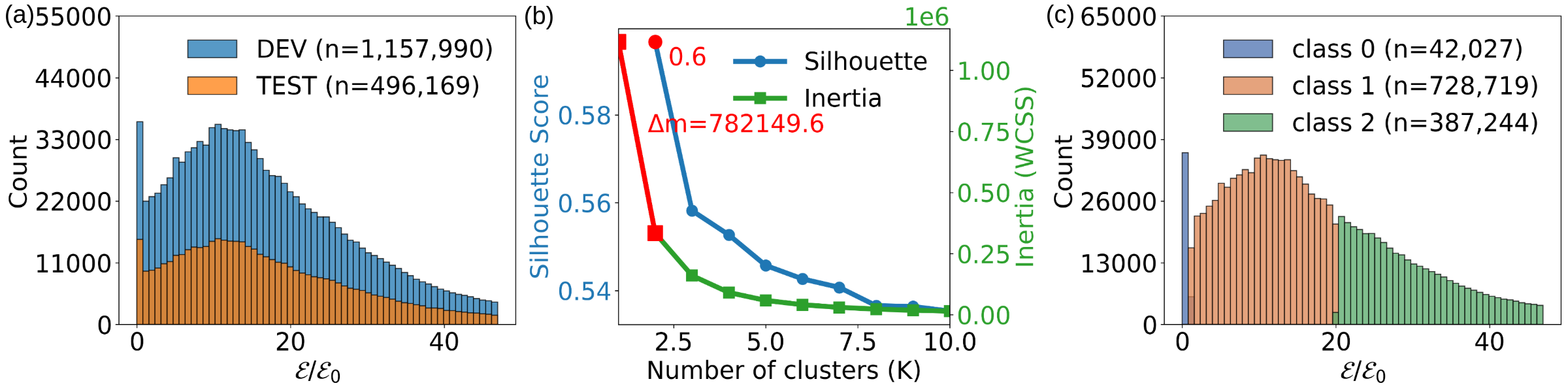}\caption{(a) Dataset divided into DEV and TEST splits; the histograms of normalized energy $\mathcal{E}/\mathcal{E}_0$ exhibit comparable right-skewed distributions. (b) On the DEV set, samples with ( $\mathcal{E}/\mathcal{E}_0$ $<$ 1) are manually assigned to class 0, while the remaining data undergo unsupervised clustering. Both the silhouette score and the within-cluster sum of squares (WCSS) identify \(K = 2\) as the optimal number of clusters, yielding two additional learned classes. (c) The resulting three-class structure,  is visualized through class-wise histograms of $\mathcal{E}/\mathcal{E}_0$.}
    \label{fig:clustering}
\end{figure*}
\section{Methodology}
Any quantum battery should be able to lead to a larger ergotropy than the classical case. The question that we address here is: which of the battery's parameters or variables can be used to identify or predict high performance batteries, i.e  with high ergotropy?  The presence of so many system parameters lead to a lot of permutations  which need to be changed to identify a high performance quantum battery. The common choice would be to focus on the role of the four energies, $\epsilon_1=\epsilon_2=\epsilon, \epsilon_a, \epsilon_b$, or the quantum parameters, $p_c,p_h,\tau$, the thermal parameters at the terminals, $T_c,T_h, T_\ell$, the thermodynamic parameters, $Q_h,Q_c, W,F, \eta$ and the set of four dimensionless cumulants $C^{(i)}, i =1,2,3,4$. It is hence an open and natural question: which of the system parameter combinations are best suited to be varied so as to identify high ergotropy batteries? Be it the battery's personal parameters (energy, coherence and temperatures), or the thermodynamic parameters or the statistical parameters, each of these quantities have non-trivial dependencies on the ergotropy along with unknown sensitivities.  In order to do so, we aim to learn  the following mappings based on the type of parameter space
\begin{align}
f_E &: \{\epsilon, \epsilon_b, \epsilon_a\} \to {\cal E}/{\cal E}_o, \\
f_T &: \{T_h, T_c, T_\ell\} \to {\cal E}/{\cal E}_o,\\
f_Q &: \{p_c, p_h, \tau\} \to {\cal E}/{\cal E}_o,\\
f_{Th} &: \{\eta, Q_c, F\} \to {\cal E}/{\cal E}_o,\\
f_C &: \{C^{(i)}\} \to {\cal E}/{\cal E}_o,
\label{eq-manual-maps}
\end{align}
and check which of the subspaces are good for learning the ergotropy using ML techniques. The full mapping, $f_{Full}:\{all~parameters\} \to {\cal E}/{\cal E}_o$ will of course  be the best, but all parameters cannot be tuned or varied simultaneously in practice. Our goal is to identify a minimal set of parameters which can used to predict the ergotropy. The variables in the above mappings will be treated as features in the language of ML.  We now proceed to outline the general ML protocol that we employ. The methodological framework is designed as a multi-stage process encompassing data partitioning, an unsupervised labeling scheme,  and a supervised classification pipeline. Each stage is constructed to rigorously prevent information leakage from  test sets into any aspect of training, labeling, or model selection.
\begin{table*}
\caption{Model families, key hyperparameters (search space), and special pipeline settings.}
\label{tab:Pipeline_settings}
\renewcommand{\arraystretch}{0.1}
\begin{ruledtabular}
\begin{tabular}{p{1cm} p{6cm} p{6cm}}
Model & Hyperparameters & Notes / Pipeline \\
\colrule

RF &
\begin{tabular}{@{}l@{}}
n\_estimators $\in \{300, 500, 800\}$ \\
max\_depth $\in \{\text{None}, 16, 24\}$ \\
min\_samples\_leaf $\in \{1, 2, 5, 10\}$ \\
max\_features $\in \{\text{"sqrt"}, 0.5\}$ \\
\end{tabular}
&
\begin{tabular}[t]{@{}l@{}}
Imputer (median) \\
Class weights = balanced \\
No scaler required \\
RandomSearchCV \\
\end{tabular} \\

\colrule

HGB &
\begin{tabular}[t]{@{}l@{}}
learning\_rate $\in \{0.05, 0.1\}$ \\
max\_depth $\in \{\text{None}, 6, 10\}$ \\
max\_leaf\_nodes $\in \{31, 63, 127\}$ \\
min\_samples\_leaf $\in \{20, 50, 100\}$ \\
l2\_regularization $\in \{0, 10^{-3}, 10^{-2}\}$ \\
\end{tabular}
&
\begin{tabular}[t]{@{}l@{}}
Log-loss \\
Validation fraction = 10\% \\
Max bins = 255 \\
Class weights = balanced \\
BayesSearchCV \\
\end{tabular} \\

\colrule

MLP &
\begin{tabular}[t]{@{}l@{}}
hidden\_layer\_sizes $\in \{(128), (128,64), (256,128)\}$ \\
activation $\in \{\text{relu}, \text{tanh}\}$ \\
alpha $\in \{10^{-4}, 3\times10^{-4}, 10^{-3}\}$ \\
learning\_rate\_init $\in \{10^{-3}, 3\times10^{-4}\}$ \\
\end{tabular}
&
\begin{tabular}[t]{@{}l@{}}
Imputer (median) + StandardScaler \\
Early stopping (max\_iter=1000, \\ n\_iter\_no\_change=20) \\
Batch size = 128 \\
RandomSearchCV \\
Solver = Adam \\
\end{tabular} \\

\end{tabular}
\end{ruledtabular}
\end{table*}
\begin{figure*}
 \centering
        \includegraphics[width=\textwidth]{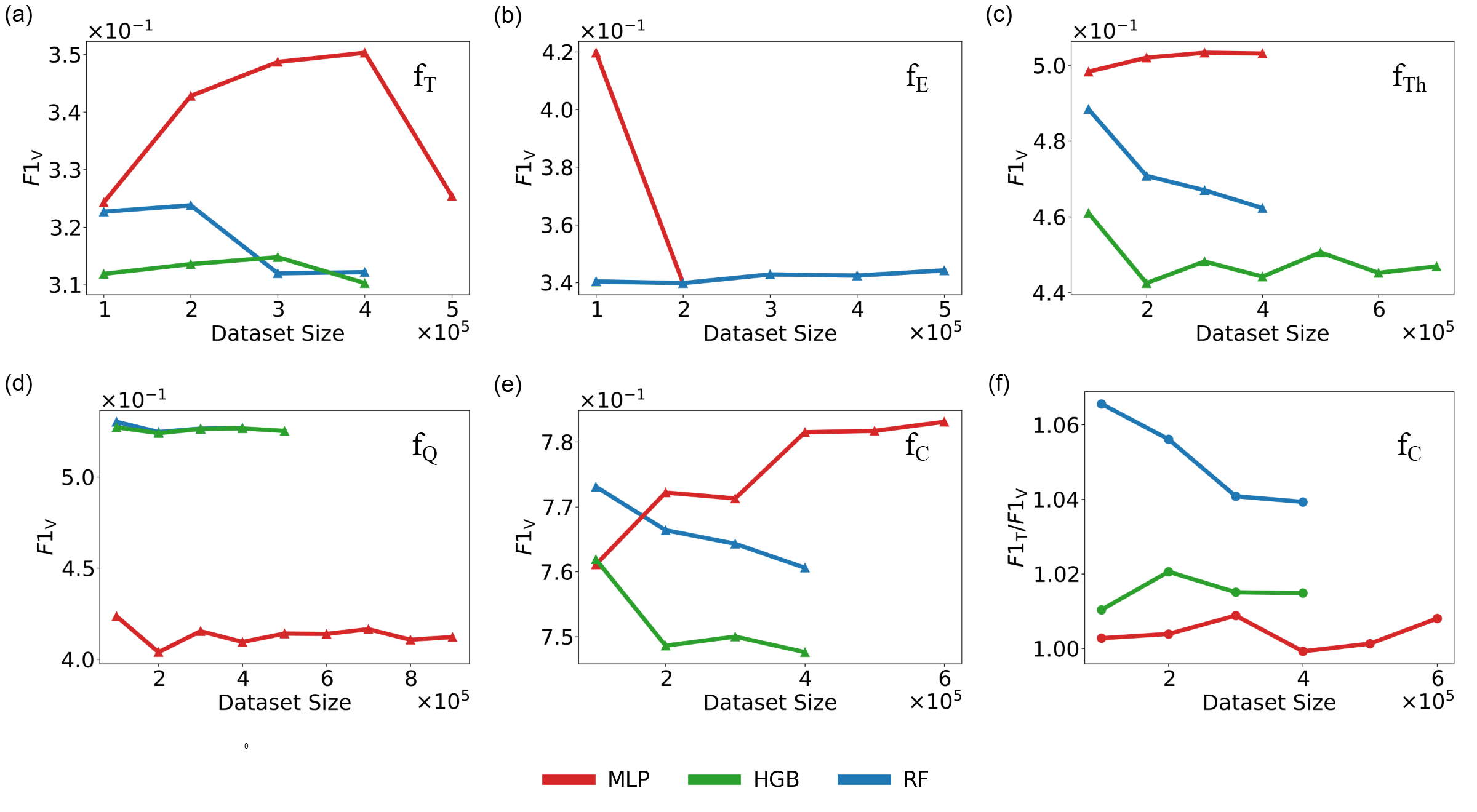}\caption{Scaling of performance with dataset size for five parameter mappings. Panels (a-e) show the validation macro-F1, \(F1_V\), versus dataset size for \(f_T\), \(f_E\), \(f_{Th}\), \(f_Q\), and \(f_C\); panel (f) shows the train-to-validation ratio \(F1_T/F1_V\) for \(f_C\) to diagnose overfitting. Curves compare MLP (red), HistGradientBoosting (green), and Random Forest (blue). The x-axis in all panels is dataset size (\(\times 10^{5}\)).
\label{fig:initial_mapping}}
\end{figure*}
\subsection{Synthetic Data Generation}

We begin by generating the dataset required for the learning process through random sampling of the independent parameters within the ranges listed in Table~\ref{tab:MCintervals} and plugging each individual value or battery configuration in Eq.(\ref{Louv-eq}). 
The sampling method employs independent draws for the ground state energy $\epsilon \in [0.01, 2]$ supplemented by random gaps $\delta_1, \delta_2 \sim (0.01, 2)$ such that $\epsilon_b = \epsilon + \delta_1$ and $\epsilon_a = \epsilon_b + \delta_2$ to enforce the constraint $\epsilon < \epsilon_b < \epsilon_a$. Through nested loops across six independent features $( p_c, p_h, T_c, T_h, T_{\ell}, \tau)$ in combination with the three pre-sampled energy triplets $(\epsilon, \epsilon_b, \epsilon_a)$, we generate $8^7$ distinct quantum battery configurations for the nine parameters. Pseudo random seeds were used to ensure reproducibility across all calculations. Requiring over two million eigenvalue evaluations, this systematic parameter space exploration ensures an exhaustive sampling of the quantum battery's operational regimes.
For each sampled parameter tuple $(T_c,T_h,T_{\ell},p_c,p_h,\tau,\epsilon,\varepsilon_b,\varepsilon_a)$ we then compute the
steadystate populations and coherence by solving the microscopic master equation, Eq. (\ref{Louv-eq}). For each sampled parameter tuple, we compute the complete set of thermodynamic $(W, F, Q_h, Q_c, \eta)$.  Likewise, we identify the passive states  by analysing the density matrix elements for each tuple and evaluate $\mathcal{E}$ and $\mathcal{E}_0$.  To evaluate the cumulants, $j^{(i)}$ and $j^{(i)}_0$, we numerically evaluate the eigenvalues of Eq. (\ref{Louv-eq}) for each tuple as a function of the counting field $\lambda$ and use Eq. (\ref{eq-cum}). The scaled ratios $C^{(i)} = j^{(i)}/j^{(i)}_0$ $(i=1,2,3,4)$ are then evaluated. 
The resulting multidimensional  dataframe consisting of the nine parameters, thermodynamic variables, ergotropy and cumulants form the initial feature space for our machine learning analysis. There are 16 such features, $( T_c, T_h, T_{\ell}, \tau, p_c, p_h, \epsilon , \epsilon_b, \epsilon_a, F, Q_c, \eta = W/Q_h, C^{(1)}, C^{(2)}, C^{(3)}, C^{(4)})$.


The raw dataframe is next subjected to systematic filtering to ensure physical consistency and numerical stability. We remove data arising from zero-valued or near zero-valued numerical errors and eliminate any degenerate configurations in which quantum effects vanish completely from the raw dataset.  This comprehensive pipeline reduces the dataset from 2,085,878 to 1,654,159 samples. 
\subsection{Unsupervised Classification of Ergotropy}
Firstly, we aim to identify the different types of ergotropy present in the 1,654,159 samples of the generated data. The objective is to identify separate classes (which we denote as $y$, $y\in\{0,\dots,K\}$) in the ergotropy. The optimal value of $K$ is determined by an unsupervised clustering technique  as follows.  The dataframe is created by constructing the space ${\cal D}:=\{(\mathbf{x}_i, r_i)\}_{i=1}^N, N = 1,654,159$, where $\mathbf{x}_i$ is the feature space (belonging to the  16 variables) and $r_i={({\cal E}/{\cal E}_o)}$ is a one dimensional space. An this stage, samples are partitioned once into a development set (DEV, 70\%) and a test set (TEST, 30\%) before any learning or classification-labeling. The data is shown in Fig (\ref{fig:clustering}a). The partitioning is based on  a group-aware split \cite{Varoquaux2017CV,Kaufman2012Leakage} with keys $[{p_h,T_c,T_h,T_l,\tau,p_c}]$ to avoid group leakage, ensuring that samples from the same experimental configuration do not appear in both sets. We excluded the energies from the keys since it allowed the latter. The TEST data are kept separated and are used only for manual comparison post prediction. 

\begin{table*}
\caption{TEST-set performance metrics by mapping.}
\label{tab:Initial_6_mapping}
\renewcommand{\arraystretch}{1.1}
\small
\begin{ruledtabular}
\begin{tabular}{l c c c r r r r r}
Mapping & Model & Variant & Size & Accuracy& F1 & MCC & ROC & PR \\
\colrule
$f_T:\{T_h,T_c,T_\ell\}$ 
& RF  & F1  & 3L & 0.3666 & 0.3140 & 0.1445 & 0.6344 & 0.3998 \\
& HGB & MCC & 2L & 0.3682 & 0.3161 & 0.1463 & 0.6422 & 0.4031 \\
& MLP & F1  & 4L & 0.6376 & 0.3510 & 0.1382 & 0.6770 & 0.4313 \\
\colrule
$f_E:\{\epsilon,\epsilon_b,\epsilon_a\}$ 
& RF  & F1  & 1L & 0.4037 & 0.3419 & 0.2132 & 0.7312 & 0.4789 \\
& HGB & F1  & 2L & 0.4037 & 0.3419 & 0.2132 & 0.7312 & 0.4789 \\
& MLP & F1  & 1L & 0.5283 & 0.4222 & 0.2749 & 0.7312 & 0.4789 \\
\colrule
$f_{Th}:\{\eta,Q_c,F\}$ 
& RF  & MCC & 1L & 0.5909 & 0.4904 & 0.3935 & 0.8080 & 0.5409 \\
& HGB & MCC & 5L & 0.5197 & 0.4517 & 0.3727 & 0.8228 & 0.5511 \\
& MLP & F1  & 4L & 0.7555 & 0.5041 & 0.4961 & 0.8398 & 0.5668 \\
\colrule
$f_Q:\{p_c,p_h,\tau\}$ 
& RF  & F1  & 4L & 0.5668 & 0.5243 & 0.3143 & 0.7758 & 0.5445 \\
& HGB & MCC & 3L & 0.5636 & 0.5220 & 0.3157 & 0.7766 & 0.5457 \\
& MLP & MCC & 3L & 0.6603 & 0.4484 & 0.2658 & 0.7964 & 0.5552 \\
\colrule
$f_C:\{C^{(i)}\}$ 
& RF  & MCC & 4L & 0.8252 & 0.7610 & 0.6682 & 0.9470 & 0.8632 \\
& HGB & F1  & 1L & 0.8182 & 0.7532 & 0.6592 & 0.9425 & 0.8534 \\
& MLP & MCC & 6L & 0.8494 & 0.7925 & 0.6889 & 0.9493 & 0.8619 \\
\colrule
$f_{Full}:\{\text{all parameters}\}$ 
& RF  & F1  & 6L & 0.9940 & 0.9913 & 0.9877 & 0.9999 & 0.9997 \\
& HGB & F1  & 4L & 0.9937 & 0.9894 & 0.9871 & 0.9999 & 0.9996 \\
& MLP & MCC & 3L & 0.9954 & 0.9925 & 0.9905 & 0.9999 & 0.9998 \\
\end{tabular}
\end{ruledtabular}
\end{table*}

\begin{figure*}
    \centering
    \includegraphics[width=\textwidth]{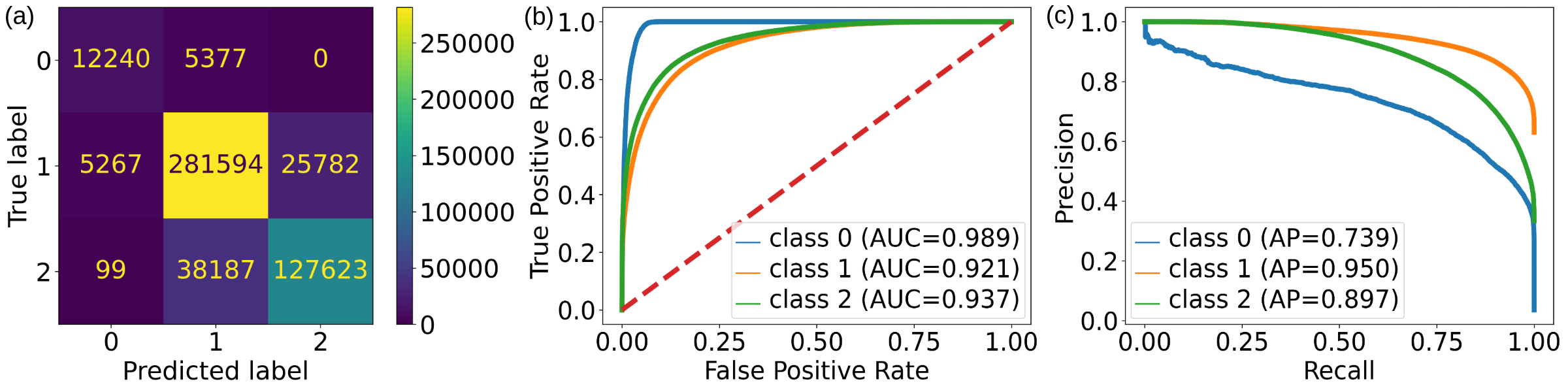}
    \caption{Cumulants mapping \(f_C:\{C^{(i)}\}\rightarrow \mathcal{E}/\mathcal{E}_0\) on the TEST set for the MLP model. (a) Confusion matrix, (b) one-vs-rest ROC curves (per-class AUCs shown), and (c) precision-recall (PR) curves (per-class APs shown). The three classes correspond to ergotropy regimes: class 0, \( \mathcal{E}/\mathcal{E}_0 < 1 \); class 1, mid; class 2, high. The model shows strong separability with most residual confusion between classes 1 and 2, yielding macro ROC-AUC \(\approx 0.921\text{--}0.989\) and macro PR-AUC \(\approx 0.739\text{--}0.950\).
    \label{fig:cumulants}}
\end{figure*}

Before, we begin any learning on this DEV dataset, we label a class manually. We do so by defining a rule-based class ($y=0$) and reserve it for a low-ergotropy regime, $
({\cal E}/{\cal E}_o) < 1 \equiv r_i < 1$. All remaining DEV points ($r_i\ge 1$) are clustered based on the standardized $r_i$ values using $\mathrm{StandardScaler}(r)$  and KMeans algorithm to discover additional classes ($y\in\{1,\dots,10\}$). We compute the silhouette score for $r_i>1$,
\begin{equation}
\label{eq:silhouette}
\mathrm{sil}(K) \;=\; \frac{1}{|\mathcal{D}_{\text{DEV}}|}\sum_{i\in \mathcal{D}_{\text{DEV}}}\frac{b(i)-a(i)}{\max\{a(i),b(i)\}}\,,
\end{equation}
\textit{where} $K$ is the candidate number of clusters. $\mathcal{D}_{\text{DEV}}$ is the DEV split used for unsupervised fitting (with $r_i\ge1$); $a(i)$ is the mean distance from sample $i$ to all other points in its own cluster (intra-cluster distance); $b(i)$ is the smallest mean distance from $i$ to points in any other cluster (best alternative cluster). 
The within-cluster sum of squares (WCSS) is
\begin{equation}
\label{eq:wcss}
\mathrm{WCSS}(K)
= \sum_{k=1}^{K} \sum_{i \in \mathcal{C}_k} \left\lVert z_i - \mu_k \right\rVert^2 ,
\end{equation}
\textit{where} $\mathcal{C}_k$ is the index set of points assigned to cluster $k$; $z_i=\mathrm{StandardScaler}(r_i)$ is the standardized scalar used for KMeans in this 1D pre-clustering; $\mu_k$ is the centroid (mean) of cluster $k$ in the $z$-space.
$K$ is then determined, considering both the inertia curve and the silhouette score as shown in Fig (\ref{fig:clustering}b). When both metrics were available, the silhouette-optimal 
$K$ was prioritised.  If the silhouette score did not yield a clear maximum, the inertia-based elbow point was used instead. We find the optimal value, $K = 2$ for $r_i\ge 1$. Hence, we obtain a total of three (one manual, two predicted) different ergotropy classes which represent three different physical regimes  as shown in  Fig (\ref{fig:clustering}c). Class 0 represents a low ergotropy regime. Class 1 represents a moderately high ergotropy regime and class 2 represents a high ergotropy regime. The word high is used to indicate coherence-enhanced values of the ergotropy. In the low ergotropy regime, coherences diminish the ergotropy. In the other two regimes, coherences amplify the ergotropy with the second class being the most enhanced. Our next objective is to establish which minimal set of the 16 parameters are best suited to predict the ergotropy into these three classes by considering the mappings defined at the beginning of the section, Eq. (9-\ref{eq-manual-maps}). We numerically aim to establish these mappings as a supervised classification.

\subsection{ Supervised Classification of Parameters into Predicted Ergotropy Classes}
We now proceed to learn the following classification, $f_E: \{ \epsilon,\epsilon_b,\epsilon_a\}\to {\cal E}/{\cal E}_o$,  $f_T: \{ T_h,T_c, T_\ell\}\to {\cal E}/{\cal E}_o$,  $f_Q: \{p_c,p_h,\tau\}\to {\cal E}/{\cal E}_o$,  $f_{Th}: \{ \eta,Q_c,F\}\to {\cal E}/{\cal E}_o$ and  $f_C: \{ C^{(i)}\}\to {\cal E}/{\cal E}_o$, where $ {\cal E}/{\cal E}_o$ has already been separated and labeled into three classes. The developmental dataframe for each mapping is constructed separately. For each of the six classifications, the six different dataframes consist of only the parameters mentioned in the argument for the respective mapping. The input vector in each dataframe consists of the battery-parameters while the output vector is the ergotropy-ratio values labeled as classes.  Since, such classifications are now commonly used, we skip the other technical details \cite{Rumelhart1986Backprop,Friedman2001GBM,Breiman2001RandomForests} related to the supervised methodology.  Three supervised model families were considered, Random Forest (RF), HistGradientBoosting (HGB), and MLP (Multi Layer Perceptron) classifiers. All classifications were performed using the standard python's sklearn library using a 3-fold SGK to label training and validation sets on the DEV data. Preliminary tests showed these outperformed other models based on  Naive Bayes, AdaBoost and Logistic Regressions. Hyperparameter search spaces and other relevant notes are reported in Table~\ref{tab:Pipeline_settings}.

In each manually considered mapping, Eq. (9-\ref{eq-manual-maps}) for each model family, two independent searches were performed at different data sizes.  One search was optimized with the performance metric macro-F1  score and  the other used the Matthew's Correlation Coeefficient, MCC. The validation F1 score ($F_{1v}$) along with the ratio between the training and validation F1 score ($F_{1T}/F_{1v}$) as a function of the discretized DEV dataset size for the three models is shown in Fig. (\ref{fig:initial_mapping}a-h). The performance with the TEST data (along with other commonly used performance metrics) are provided in  Table \ref{tab:Initial_6_mapping} and  performance for TEST data particularly for the cumulants mapping is shown in Fig. (\ref{fig:cumulants}).
The confusion matrix provides a direct visualization of how well each ergotropy class is predicted, revealing specific misclassifications between low-, mid-, and high-ergotropy regimes.
The ROC curve AUC$_{ROC}$ measures the model's ability to distinguish between low-, mid-, and high-ergotropy regimes, with higher values indicating sharper class separability. Since the dataset is moderately imbalanced, the Precision-Recall curve AUC$_{PR}$ is also included, as it better reflects performance on minority (high-ergotropy) classes. Reporting both metrics provides a balanced view ROC captures overall separability, while PR highlights precision under imbalance.

\begin{table}[t]
\caption{Classifier models and fixed hyperparameters for feature importance analysis.}
\label{tab:feat_ananlysis_hyperparameters}
\renewcommand{\arraystretch}{1.2}
\begin{ruledtabular}
\begin{tabular}{ll}
\textbf{Model} & ~~~~~\textbf{Hyperparameters} \\
\colrule
Random Forest &
\parbox[t]{6cm}{%
$n_{\mathrm{estimators}}=100$ \\[2pt]
$\mathrm{max\_depth}=10$%
} \\
Gradient Boosting &
\parbox[t]{6cm}{%
$n_{\mathrm{estimators}}=100$ \\[2pt]
$\mathrm{learning\_rate}=0.1$ \\[2pt]
$\mathrm{max\_depth}=5$%
} \\
MLP &
\parbox[t]{6cm}{%
Hidden Layers: $(100,50)$ \\[2pt]
Activation: ReLU \\[2pt]
Solver: Adam \\[2pt]
$\mathrm{max\_iter}=500$%
} \\
\end{tabular}
\end{ruledtabular}
\end{table}


We next aim to establish the importance of the 18 battery features on prediction of the ergotropy using a leakage-free protocol. Theoretically, the mapping that we want to identify is $f:\{ \mathbf{x}_i\}_{\min}\to {\cal E}/{\cal E}_0$, where $\{ \mathbf{x}_i\}_{\min}$ is the minimal set of input features required for high prediction accuracy of the ergotropy classes from the list of mappings as shown in Table \ref{tab:featuremappingslist}. We use the RF, GB and MLP classifier with fixed hyperparameters with a 7:3 split between  training  and validation and is shown in Table \ref{tab:feat_ananlysis_hyperparameters}.  All pipelines included median imputation while the MLP included feature standardization. Feature importance was measured by the drop in model performance (decrease in macro-F1 ($F1_{\mathrm{macro}}$ value during validation) when that feature's values were randomly shuffled across 5 trials. The results are compiled in Fig. (\ref{fig:featimpo}).

\begin{table}
\caption{Feature subspaces ordered by complexity.}
\label{tab:featuremappingslist}
\renewcommand{\arraystretch}{1.1}
\begin{ruledtabular}
\begin{tabular}{cl}
\textbf{Mapping ID} & \textbf{Feature Set} \\
\colrule
$f_1$ & $\{C^{(4)},\,p_c\}$ \\
$f_2$ & $\{C^{(4)},\,p_c,\,T_h\}$ \\
$f_3$ & $\{C^{(4)},\,p_c,\,Q_h\}$ \\
$f_4$ & $\{C^{(4)},\,p_c,\,Q_h,\,\epsilon_a\}$ \\
$f_5$ & $\{C^{(4)},\,p_c,\,Q_h,\,T_h\}$ \\
$f_6$ & $\{C^{(4)},\,p_c,\,Q_h,\,T_h,\,T_\ell\}$ \\
$f_7$ & $\{C^{(4)},\,p_c,\,T_h,\,T_\ell,\,Q_h,\,Q_c,\,\epsilon_a,\,\epsilon_b\}$ \\
\end{tabular}
\end{ruledtabular}
\end{table}

\section{Results and Discussion}

We already discussed Figure~\ref{fig:clustering}, which established three distinct classes for the ergotropy to be used for the subsequent supervised learning. From Fig. (\ref{fig:initial_mapping}a-d), based on the relatively low $F_{1v}$ scores at different data sizes for the three models (RF, HGB, MLP), we conclude that the mappings $f_T, f_E, f_{Th}$ and $f_Q$ are not suitable for ergotropy prediction. Hence, one should not fiddle around with the thermal ($T_c,T_h, T_\ell$), the energies ($\epsilon_i,i=a,b,1,2$), thermodynamic variables ($\eta,F,Q_c$) or quantum parameters ($p_c,p_h,\tau$), separately, in the quest for identifying high ergotropy batteries. Even though these parameters are quite popular, these cannot be used to predict ergotropy in quantum batteries. From the values of the performance indicators in Fig. (\ref{fig:initial_mapping}e,f), we learn that the mapping $f_C$ is relatively better than the others with  validation F1-scores of $~$78\% within the 0.4-0.6 million data range for the MLP model. The other two models learnt to classify correctly, albeit with slightly lower validation scores. Using the MLP classifier, we manually compare the prediction data with the initially seperated TEST data. The confusion matrix, ROC and precision vs recall curves are shown in Fig. (\ref{fig:cumulants}a-c) for the TEST data. The accuracy, F1 score, MCC, ROC amd PR are respectively 0.8494, 0.7925, 0.6889, 0.9493 and 0.8619. This result strongly suggests that the statistical fluctuations of the system, captured by the cumulants, contain additional information about the ergotropy that is not present in the battery's lone parameters or thermodynamic variables alone. We can conclude that knowledge and control over the cumulant space can aid in the identification of high ergotropy batteries or better batteries, more than the actual battery parameters. Note that,  a model trained on all available parameters  achieved near-perfect scores (MCC $>$ 0.99). However, the objective is to achieve good prediction with as few parameters as possible.

\begin{figure*}
    \centering
    \includegraphics[width=\textwidth]{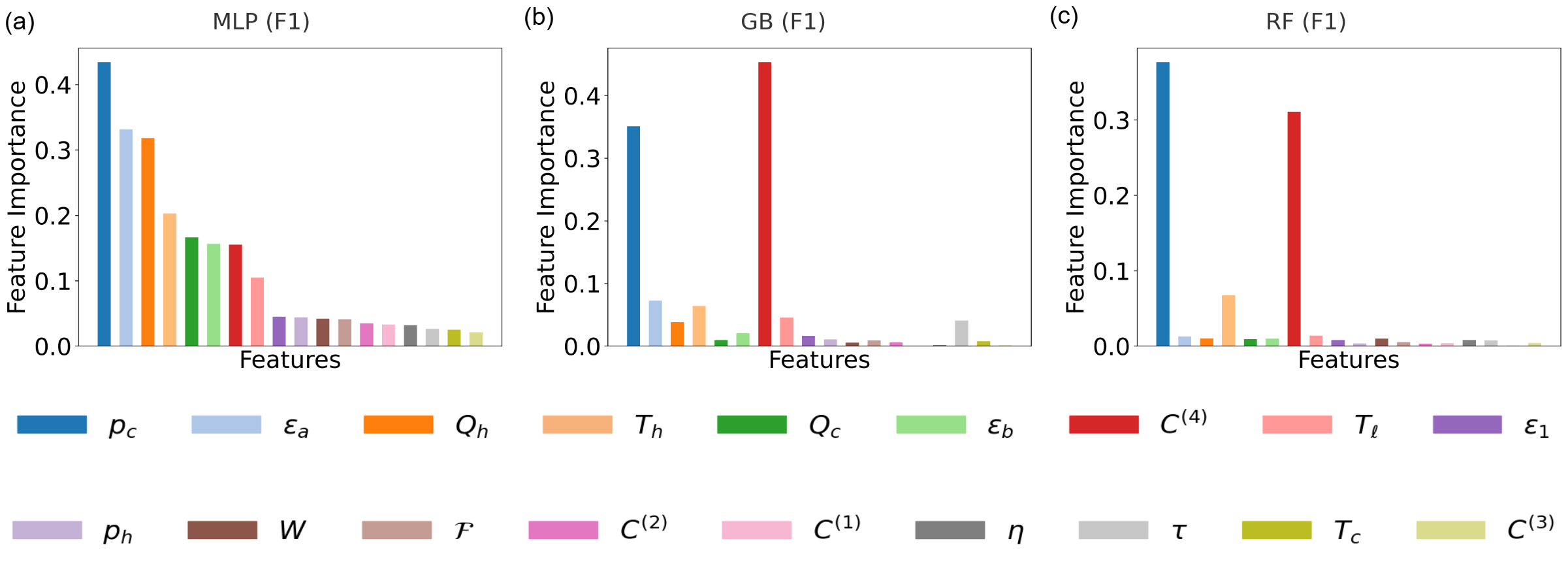}
    \caption{F1-based feature-importance comparison across three models multi layer perceptron (MLP), HistGradientBoosting (HGB), and Random Forest (RF). Bars correspond to the physical variables \(\{p_c,\ \varepsilon_a,\ Q_h,\ T_h,\ Q_c,\ \varepsilon_b,\ C^{(4)},\ T_\ell,\ \varepsilon_1,\ p_h,\ W, F, \ C^{(2)},\ C^{(1)},\ \eta,\ \tau,\ T_c,\ C^{(3)}\}\). Consistently high importances of \(p_c\), and \(C^{(4)}\) indicate their dominant role in determining classification performance across all models.}
    \label{fig:featimpo}
\end{figure*}

\begin{figure*}
 \centering
        \includegraphics[width=\textwidth]{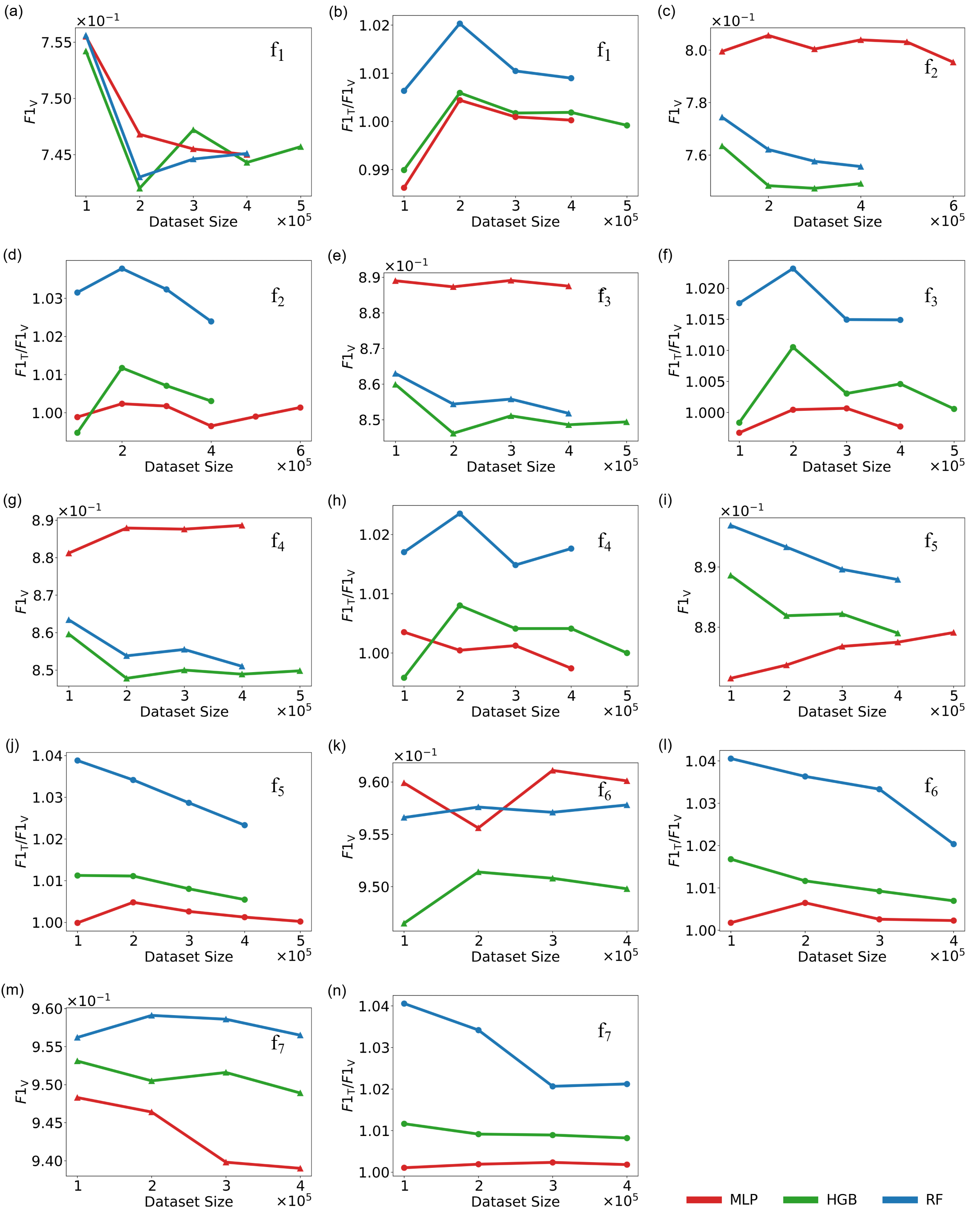}\caption{Performance scaling with dataset size for seven feature mappings \(f_{1}\)-\(f_{7}\) (see Table~\ref{tab:Feat_engg_mapping}). Panels are paired by mapping:
\((a,b)\,f_{1}\), \((c,d)\,f_{2}\), \((e,f)\,f_{3}\), \((g,h)\,f_{4}\), \((i,j)\,f_{5}\), \((k,l)\,f_{6}\), \((m,n)\,f_{7}\).
Panels \((a,c,e,g,i,k,m)\) show the validation macro-F1 \(F1_V\) versus dataset size; Panels \((b,d,f,h,j,l,n)\) show the train-to-validation ratio \(F1_T/F1_V\) versus dataset size. Scientific-notation ticks indicate scale where shown. In all panels, the x-axis is dataset size (\(\times 10^{5}\) where indicated), and curves compare MLP (red), HistGradientBoosting (green), and Random Forest (blue).
\label{fig:featimpo_mapping}}
\end{figure*}

\begin{figure*}
    \centering
    \includegraphics[width=\textwidth]{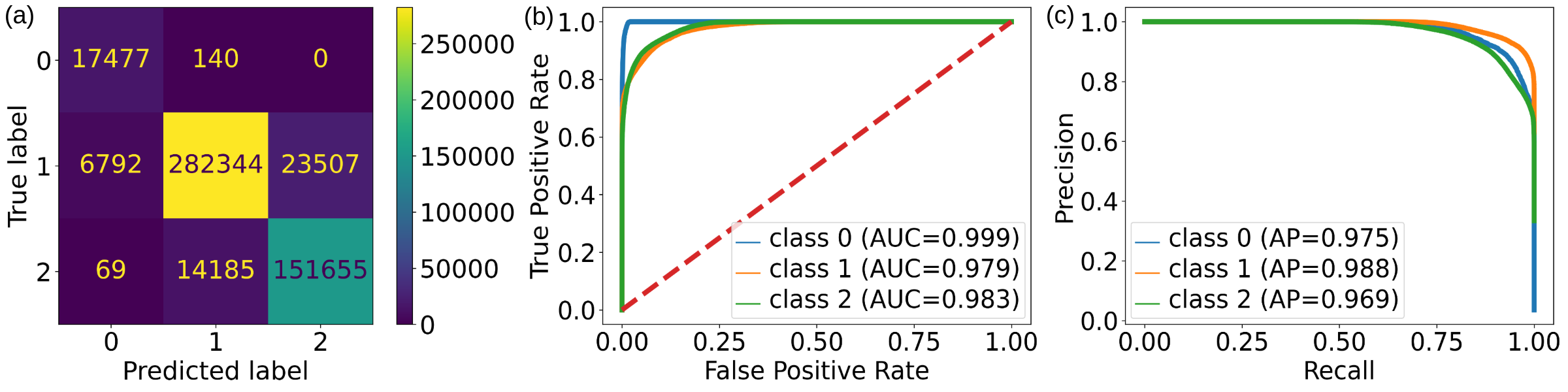}
    \caption{Feature set \(f_{5}:\{C^{(4)},\,T_h,\,Q_h,\,p_c\}\) on the TEST split with the MLP model. Panels: (a) confusion matrix, (b) one-vs-rest ROC curves with per-class AUCs, and (c) precision-recall curves with per-class average precision (AP). Classes denote ergotropy regimes: 0, \(\mathcal{E}/\mathcal{E}_0<1\); 1, mid; 2, high. The model achieves near-perfect discrimination the per-class ROC-AUCs are \(\approx 0.999,\,0.979,\,0.983\) and APs are \(\approx 0.975,\,0.988,\,0.969\) with only minor residual confusion, primarily between classes 1 and 2.
    \label{fig:featmapping_5}}
\end{figure*}

\begin{figure*}
    \centering
    \includegraphics[width=\textwidth]{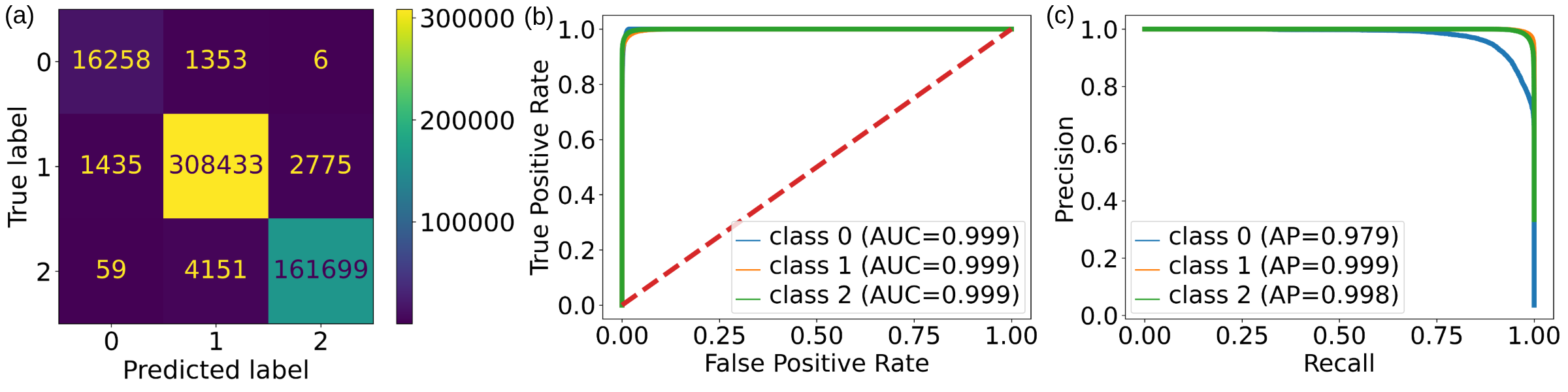}
    \caption{Feature set \(f_{6}=\{C^{(4)},\,T_h,\,Q_h,\,p_c,\,T_\ell\}\) on the TEST split with the MLP model. Panels: (a) confusion matrix, (b) one-vs-rest ROC curves with per-class AUCs, and (c) precision-recall curves with per-class average precision (AP). Classes denote ergotropy regimes: 0, \(\mathcal{E}/\mathcal{E}_0<1\); 1, mid; 2, high. The model achieves near-saturated performance per-class ROC-AUCs \(\approx 0.999\) and APs \(\approx 0.979\text{--}0.999\) with only negligible residual confusion, chiefly between classes 0 and 1.}
    \label{fig:featmapping_6}
\end{figure*}

Given the strong performance of the cumulant mapping, we now explore the possibility of creating a high-performance predictive model with machine-predicted features. We use mutual information and feature importance analysis to identify the most predictive variables using three algorithms. From the results shown in Fig. (\ref{fig:featimpo}), we find that some features or parameters outshine other parameters in classifying the ergotropy correctly. As per the feature importance score seen in Fig. (\ref{fig:featimpo}a), the MLP predicted that there are 8 system features that are relatively more important than the others. While, in Fig. (\ref{fig:featimpo}b,c), we see the fourth cumulant ($C^{(4)}$) and the coherence in the leak ($p_c$) to be the two most important features by GB and RF learning models. By analysing these feature importances, we identify a total of 7 new mappings between features and ergotropy which are listed in Table \ref{tab:featuremappingslist}. In these newly constituted mappings $f_n, n =1\ldots 7$, we train the three models into learning these mappings similar to the standard pipeline used  before.  The validation and training is to validation ratio of F1 scores  are displayed in Fig. (\ref{fig:featimpo_mapping}a-n). As shown in Fig.(\ref{fig:featimpo_mapping}a-d), a model trained on just  two features, i.e $f_1$ achieved an $F_{1v}$  of approximately 0.745 which is lower than that of the manually identified mapping $f_C$. Adding a third feature, eg. the  input or charging energy ($Q_h$) to this pair boosted the $F_{1v}$ of the mapping $f_2$  to over 0.80 as shown in Fig. (\ref{fig:featimpo_mapping}c,d). As the number of input features increase the prediction becomes better and better as seen from the scores in Fig. (\ref{fig:featimpo_mapping}e-n).  The performance of the feature-importance based mappings is a substantial improvement over the individual manually identified mapping, $f_C$, even with three or four features.  The four feature mapping, $f_5$ with the set $\{C^{(4)}, T_h, Q_h, p_c\}$  has a significantly higher MCC of $0.8226$ as tabulated on Table \ref{tab:Feat_engg_mapping}. The confusion matrices, ROC and PR curves obtained from the TEST data (MLP model) for the mapping $f_5$ are shown in Fig. (\ref{fig:featmapping_5}). The minimal feature set that performs the best in the TEST dataset is found to be $\{C^{(4)}, p_c, T_h, Q_h, T_\ell\}$, i.e, $f_6$ with an MCC = 0.96, indicating that these parameters are excellent for predicting the three ergotropy classes. The confusion matrices, ROC and PR curves for the mapping $f_6$ obtained from the MLP TEST set are shown in Fig. (\ref{fig:featmapping_6}).  Their performance is further evidenced by the highly diagonal confusion matrices, which indicate a low rate of misclassification, and the consistently high Area Under the Curve (AUC) and Average Precision (AP) scores for the ROC and PR curves, respectively, with most values exceeding 0.97.  Remarkably, this level of performance is essentially identical to that of the more extensive seven-feature mapping $f_{7}$, with an MLP TEST set MCC of $0.95$. The other performance indicators, evaluated for the TEST data for each machine-predicted features based mapping, are tabulated in Table \ref{tab:Feat_engg_mapping}. We cross-validated these results with a powerful Feature-Tokenizer-Transformer (FTT)  model \cite{gorishniy2021revisiting}, which is known to be good for tabulated data \cite{hollmann2025accurate} and it further strengthened our inferences. As an instance, we remove $p_c$ from  the mapping $f_6$ and consider another four parameter mapping, $f_{8}$ composed of the set $ \{C^{(4)},T_\ell,T_h,Q_h\}$, from which we again predict the ergotropy. We  observed a drop in MCC by 19\% compared to $f_6$ with  $3\times10^{5}$ amount of data. This establishes the importance of the parameter $p_c$ between multiple models. Further, using the FTT model on the mapping $f_5$, the MCC was found to be 0.84, a minute increase over the RF, HGB and MLP calculations. 



\begin{figure*}
    \centering
    \includegraphics[width=\textwidth]{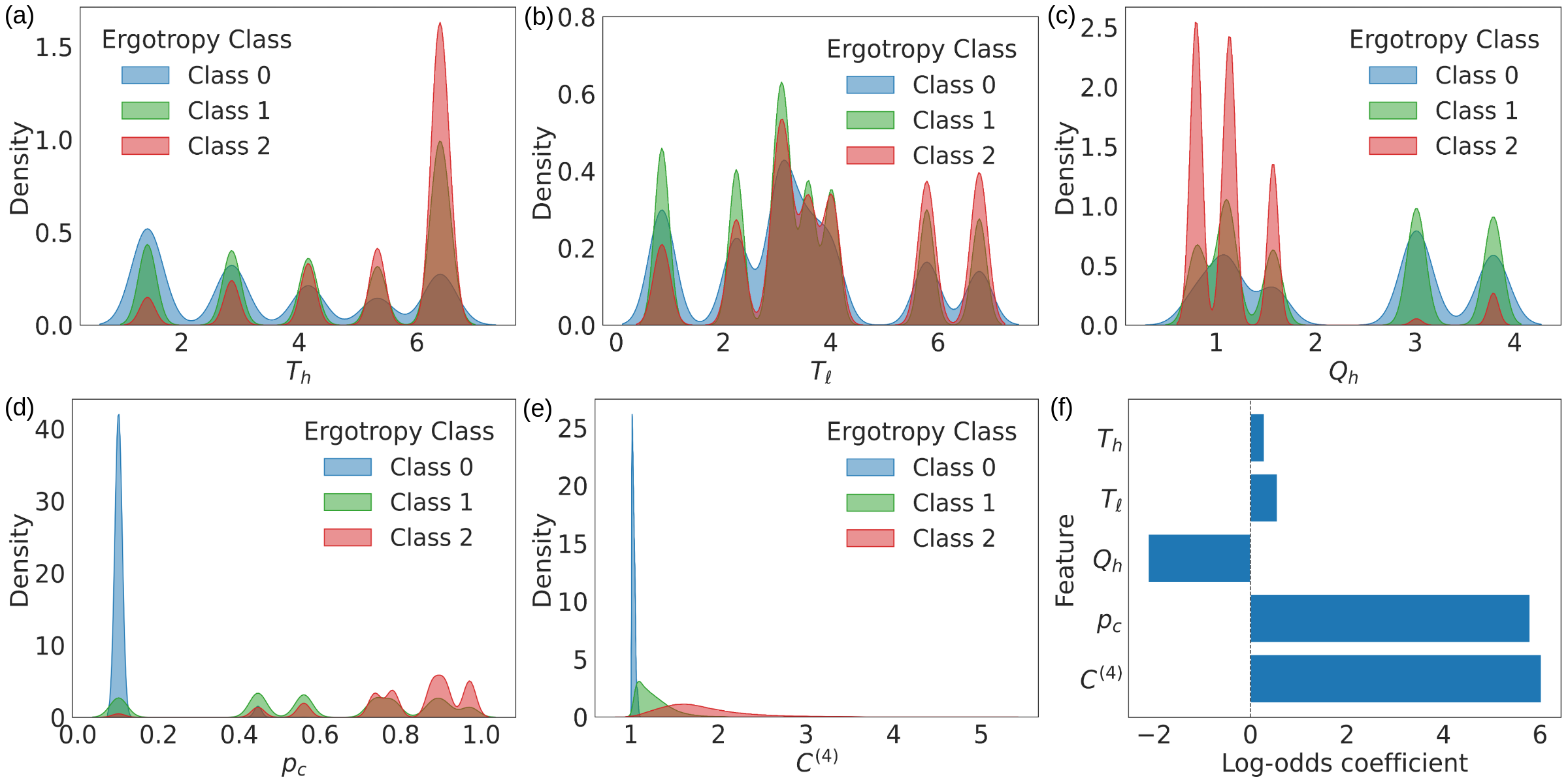}
    \caption{Feature-wise class distributions and logistic regression coefficients for identifying the high-ergotropy regime. Panels (a-e) show kernel density estimates of \(T_h\), \(T_\ell\), \(Q_h\), \(p_c\), and \(C^{(4)}\) across ergotropy classes (0: low, 1: intermediate, 2: high), highlighting the separation achieved by each parameter particularly the right-shift of \(p_c\) and \(C^{(4)}\) for class~2. Panel (f) reports log-odds coefficients from a binary logistic regression (\(y{=}2\) vs.\ \(\{0,1\}\)), indicating \(C^{(4)}\) and \(p_c\) as the strongest positive predictors, a negative contribution from \(Q_h\), and weaker positive effects from \(T_h\) and \(T_\ell\).
    \label{fig:LR}}
\end{figure*}

\begin{table*}
\caption{TEST-set performance metrics for feature subsets.}
\label{tab:Feat_engg_mapping}
\renewcommand{\arraystretch}{1.1}
\small
\begin{ruledtabular}
\begin{tabular}{l c c c r r r r r}
Mapping & Model & Variant & Size & Accuracy& F1 & MCC & ROC & PR \\
\colrule
$f_1:\{C^{(4)}, p_c\}$ 
& RF  & F1  & 3L & 0.7927 & 0.7436 & 0.6090 & 0.9278 & 0.8778 \\
& HGB & MCC & 3L & 0.7956 & 0.7458 & 0.6153 & 0.9309 & 0.8805 \\
& MLP & F1  & 2L & 0.7953 & 0.7475 & 0.6159 & 0.9303 & 0.8807 \\
\colrule
$f_2:\{C^{(4)}, p_c, T_h\}$ 
& RF  & MCC & 3L & 0.7996 & 0.7576 & 0.6195 & 0.9306 & 0.8819 \\
& HGB & F1  & 1L & 0.8001 & 0.7544 & 0.6228 & 0.9324 & 0.8846 \\
& MLP & MCC & 2L & 0.8312 & 0.8035 & 0.6520 & 0.9372 & 0.8884 \\
\colrule
$f_3:\{C^{(4)}, p_c, Q_h\}$ 
& RF  & MCC & 3L & 0.8884 & 0.8533 & 0.7843 & 0.9788 & 0.9607 \\
& HGB & F1  & 1L & 0.8892 & 0.8552 & 0.7871 & 0.9793 & 0.9617 \\
& MLP & F1  & 3L & 0.9063 & 0.8888 & 0.8082 & 0.9813 & 0.9629 \\
\colrule
$f_4:\{C^{(4)}, p_c, Q_h, \varepsilon_a\}$ 
& RF  & MCC & 3L & 0.8883 & 0.8534 & 0.7842 & 0.9787 & 0.9607 \\
& HGB & MCC & 1L & 0.8888 & 0.8550 & 0.7868 & 0.9793 & 0.9615 \\
& MLP & F1  & 3L & 0.9067 & 0.8882 & 0.8079 & 0.9811 & 0.9624 \\
\colrule
$f_5:\{C^{(4)}, T_h, Q_h, p_c\}$ 
& RF  & MCC & 3L & 0.9099 & 0.8877 & 0.8210 & 0.9857 & 0.9756 \\
& HGB & F1  & 3L & 0.9090 & 0.8819 & 0.8207 & 0.9862 & 0.9768 \\
& MLP & MCC & 5L & 0.9099 & 0.8831 & 0.8226 & 0.9868 & 0.9773 \\
& FTT & MCC & 4L & 0.9214 & 0.9100 & 0.8390 & 0.9868 & 0.9755 \\
\colrule
$f_6:\{C^{(4)}, T_h, Q_h, p_c, T_\ell\}$ 
& RF  & F1  & 4L & 0.9780 & 0.9568 & 0.9555 & 0.9988 & 0.9935 \\
& HGB & F1  & 3L & 0.9764 & 0.9500 & 0.9527 & 0.9990 & 0.9942 \\
& MLP & F1  & 3L & 0.9803 & 0.9609 & 0.9597 & 0.9990 & 0.9990 \\
\colrule
$f_7:\{C^{(4)}, p_c, T_h, T_\ell, Q_h, Q_c, \varepsilon_a, \varepsilon_b\}$ 
& RF  & F1  & 4L & 0.9784 & 0.9570 & 0.9562 & 0.9989 & 0.9936 \\
& HGB & MCC & 3L & 0.9768 & 0.9513 & 0.9535 & 0.9990 & 0.9943 \\
& MLP & F1  & 2L & 0.9753 & 0.9446 & 0.9507 & 0.9991 & 0.9934 \\
\end{tabular}
\end{ruledtabular}
\end{table*}

To elucidate the role of each feature within the optimal minimal parameter set \(\{ C^{(4)}, T_h,  Q_h, p_c, T_\ell\}\) in determining the ergotropy regimes, we analyze their class-wise probability distributions. We use a binary logistic regression to explore their influence on the high-ergotropy regime (\(y{=}2\)). The logistic regression model relates the probability \(P(y{=}2|x)\) of a sample belonging to the high-ergotropy class to its feature vector \(x = ( C^{(4)}, T_h,  Q_h, p_c, T_\ell)\) through the log-odds equation:
\begin{equation}
\ln\!\left[\frac{P(y{=}2|x)}{1 - P(y{=}2|x)}\right]
= \beta_0 + \sum_{i=1}^{5} \beta_i x_i,
\label{eq:logit}
\end{equation}
where \(\beta_0\) is the intercept and \(\beta_i\) are the fitted coefficients representing the contribution of each feature toward the likelihood of attaining the high-ergotropy state. We show the  kernel density estimates (KDEs) in Fig.~\ref{fig:LR}(a-e) which reveal class-dependent structures. 
Panel (a) shows that higher \(T_h\) values slightly prefer distribution toward mid and high ergotropy classes. This implies that increasing the charging temperature aids population activation without destabilizing coherence which lead to higher ergotropy. 
Panel (b) indicates that the storage-mode temperature is without any preferential bias towards any ergotropy regime.
The KDE of $Q_h$ in panel (c) shows a preference. Low (high) charging energy gap prefers higher (lower) ergotropy regimes. This means that larger charging energy gaps suppress coherent work extraction. 
Panel (d) exhibits a pronounced preference of larger \(p_c\) values to be in class 2 or higher ergotropy.  So, stronger  coherence in the leak enhance the probability of sustained population inversion required for higher ergotropy. 
Panel (e) shows a similar preference  of larger \(C^{(4)}\) towards higher ergotropy. Values of 
\(C^{(4)} \approx 1.5-3\) indicate a regime of extremely high coherence-dependent kurtosis. Such high kurtosis values cannot be dominated by random thermal noise and is controlled by coherent interference effects that emerge from the asymmetric coupling of the two degenerate ground states to separate stations. In this regime, the noise no longer acts as a primary decohering agent. It instead drives correlated population transfer from the two ground manifolds, producing bursts of coherent energy exchange (non-Gaussian fluctuations) and significantly enhance extractable work through sustained population inversion.  To quantify these qualitative trends, the fitted coefficients visualized in Fig.~\ref{fig:LR}(f) provide a direct measure of each feature's predictive influence. The largest positive weights correspond to \(C^{(4)}\) (\(+6.03\)) and \(p_c\) (\(+5.79\)). Thus, high coherence-driven photon exchange kurtosis and high coherence in the leak increase the likelihood of high ergotropy. 
Moderate positive effects from \(T_h\) (\(+0.29\)) and \(T_\ell\) (\(+0.56\)) indicate that balanced temperature gradients support coherent operation without destabilization. 
In contrast, the negative coefficient of \(Q_h\) (\(-2.12\)) demonstrates that larger charging energy gaps disrupt coherence and reduce ergotropy. 
Together, these results establish a quantitative hierarchy among the control parameters, identifying \(C^{(4)}\) and \(p_c\) as the principal determinants of high-ergotropy behavior instead of  the charging energy which  govern efficient energy storage in quantum batteries.

The observation of the high importance of the two terms $C^{(4)}$, the kurtosis of energy exchange in the storage mode and $p_c$, the leakage reservoir coherence, identified individually by all the three learning models, MLP, GB and RF is completely new.  The GB and RF basically considered these to be the deciding factors during ergotropy prediction, as seen from the feature importance scores of $~0.4$ in Fig. (\ref{fig:featimpo}b,c). Nothing is currently known about the sensitivity of such parameters on any quantum battery's ergotropy in the current realm of quantum thermodynamics. Kurtosis is usually a measure of quantitative outliers or large deviations in a system's behavior. The unknown dependence of the battery performance on the kurtosis of storage mode's energy exchange highlights the plausible existence of extremely non-Gaussian nature  energy exchange distribution which may not be smooth. The dependence on the leakage coherence indicates that the exchange is during storage as well as leakage since the state $\ket{b}$ interacts with both the storage subspace and leakage subspace. So, one cannot disregard the possibility of large intermittent energy jumps during  storage indicating that energy exchanges that lead to higher ergotropy could be somewhat chaotic in the presence of coherences.  
\section{Conclusion}
We introduced a cavity mediated quantum battery based on a finite quantum system where populations and coherences are coupled. The coherences are due to assymetric coupling with noisy stations while the storage dynamics is controllable through the unimodal cavity. We show how the coherent effects can modify the battery's charging , storage, leakages and the ergotropy beyond their classical values. We highlighted why the existence of several parameters render the identification of high ergotropy regimes to be a bottleneck issue. By integrating full counting statistics with machine learning techniques, we demonstrate that the higher order fluctuation space contains significant information on learning the ergotropy by exploring the utility of feature importance analysis. We showed how to capture of the most predictive battery features with minimal data  in three different ergotropy regimes, low, high and very high.  A key takeaway from our study is that high-ergotropy batteries can be identified more reliably by focusing on the first four cumulants of the system rather than merely relying on standard parameters such as energy, quantum, thermal or thermodynamic parameters in isolation. This finding suggests that higher order fluctuations are in fact central to understanding and optimizing the performance of quantum batteries under open conditions.  The fourth cumulant or kurtosis of the quanta exchange at the storage station, combined with specific thermodynamic variables like  input charging energy and leakage-mode induced coherence or the coherence in the leak, are found to be a robust predictive framework that significantly outperforms conventional standalone thermodynamic variables like work or efficiency or simple energy landscape.
Additionally, using feature importance analysis through multiple machine learning models, we  unearthed a decisive  role of the kurtosis of  quanta exchange in the storage mode and the coherence in the leakage mode in determining high-ergotropy. High coherence in the leak and large kurtosis of quanta exchanged at the storage leads to higher ergotropy. Further, lower input charging energy also prefer higher ergotropy while increasing the input charging energy favours low ergotropy.  The existence of such non-trivial dependence predicts that high ergotropy quantum batteries could rely on chaotic or non-Gaussian energy exchange during storage. 
\bibliographystyle{unsrt}
\bibliography{references-brt}

\begin{thebibliography}{10}

\bibitem{Campaioli2024RMP}
F.~Campaioli et~al.
\newblock Colloquium: Quantum batteries.
\newblock {\em Rev. Mod. Phys.}, 96(3):031001, 2024.

\bibitem{Allahverdyan2004}
A.~E. Allahverdyan, R.~Balian, and Th.~M. Nieuwenhuizen.
\newblock Maximal work extraction from finite quantum systems.
\newblock {\em Europhys. Lett.}, 67(4):565--571, 2004.

\bibitem{Ali2024ErgotropyCapacityPRA}
Asad Ali, Saif Al-Kuwari, M.~I. Hussain, Tim Byrnes, M.~T. Rahim, J.~Q. Quach, M.~Ghominejad, and S.~Haddadi.
\newblock Ergotropy and capacity optimization in heisenberg spin-chain quantum batteries.
\newblock {\em Physical Review A}, 110(5):052404, 2024.

\bibitem{Ahmadi2024PRL_NonreciprocalQB}
B.~Ahmadi et~al.
\newblock Nonreciprocal quantum batteries.
\newblock {\em Phys. Rev. Lett.}, 132:210402, 2024.

\bibitem{alicki2013}
Robert Alicki and Mark Fannes.
\newblock Entanglement boost for extractable work from ensembles of quantum batteries.
\newblock {\em Phys. Rev. E}, 87(4):042123, 2013.

\bibitem{malavazi2025chargepreservingoperationsquantumbatteries}
André H.~A. Malavazi, Borhan Ahmadi, Paweł Horodecki, and Pedro~R. Dieguez.
\newblock Charge-preserving operations in quantum batteries, 2025.

\bibitem{Shi2022CoherenceWorkPRL}
Hai-Long Shi, Shu Ding, Qing-Kun Wan, Xiao-Hui Wang, and Wen-Li Yang.
\newblock Entanglement, coherence, and extractable work in quantum batteries.
\newblock {\em Physical Review Letters}, 129:130602, 2022.

\bibitem{Kamin2020EntanglementCoherenceCharging}
F.~H. Kamin, F.~T. Tabesh, S.~Salimi, and A.~C. Santos.
\newblock Entanglement, coherence, and charging process of quantum batteries.
\newblock {\em Physical Review E}, 102:052109, 2020.

\bibitem{Francica2020}
Gianluca Francica, Felix~C. Binder, Giulia Guarnieri, Matthew~T. Mitchison, John Goold, and Francesco Plastina.
\newblock Quantum coherence and ergotropy.
\newblock {\em Phys. Rev. Lett.}, 125(18):180603, 2020.

\bibitem{Cakmak2020}
B.~{\c C}akmak.
\newblock Ergotropy from coherences in an open quantum system.
\newblock {\em Phys. Rev. E}, 102:042111, 2020.

\bibitem{Caravelli2021}
Francesco Caravelli, Bin Yan, Luis~Pedro Garc{\'{\i}}a-Pintos, and Alioscia Hamma.
\newblock Energy storage and coherence in closed and open quantum batteries.
\newblock {\em Quantum}, 5:505, 2021.

\bibitem{Perarnau2015}
Mart\'{\i} Perarnau-Llobet, Karen~V. Hovhannisyan, Marcus Huber, Paul Skrzypczyk, Jordi Tura, and Antonio Ac\'{\i}n.
\newblock Most energetic passive states.
\newblock {\em Phys. Rev. E}, 92:042147, Oct 2015.

\bibitem{Hadipour2024}
Maryam Hadipour and Soroush Haseli.
\newblock Work extraction from quantum coherence in non-equilibrium environment.
\newblock {\em Scientific Reports}, 14(1):24876, 2024.

\bibitem{OULARABI2025131003}
A.~Oularabi, A.~{El Allati}, and K.~{El Anouz}.
\newblock Enhancing ergotropy of quantum batteries through coherence and non-markovianity.
\newblock {\em Physica A: Statistical Mechanics and its Applications}, 679:131003, 2025.

\bibitem{khoudiri2025coherencedrivenquantumbatterycharging}
Achraf Khoudiri, Abderrahman Oularabi, Khadija~El Anouz, İlkay Demir, and Abderrahim~El Allati.
\newblock Coherence-driven quantum battery charging via autonomous thermal machines: Energy transfer, memory effects, and ergotropy enhancement, 2025.

\bibitem{ahmadisuper}
Borhan Ahmadi, Pawe\l{} Mazurek, Shabir Barzanjeh, and Pawe\l{} Horodecki.
\newblock Superoptimal charging of quantum batteries via reservoir engineering: Arbitrary energy transfer unlocked.
\newblock {\em Phys. Rev. Appl.}, 23:024010, Feb 2025.

\bibitem{PhysRevResearch.2.013095}
Stefano Gherardini, Francesco Campaioli, Filippo Caruso, and Felix~C. Binder.
\newblock Stabilizing open quantum batteries by sequential measurements.
\newblock {\em Phys. Rev. Res.}, 2:013095, Jan 2020.

\bibitem{Sen2023NoisyQB}
Kornikar Sen and Ujjwal Sen.
\newblock Noisy quantum batteries.
\newblock {\em arXiv preprint arXiv:2302.07166}, 2023.

\bibitem{Wang2024PRA_COQB}
L.~Wang et~al.
\newblock Cavity-optomechanical quantum battery.
\newblock {\em Phys. Rev. A}, 110(6):062204, 2024.

\bibitem{Tirone2025NoisyQBEfficiency}
S.~Tirone et~al.
\newblock Quantum work extraction efficiency for noisy quantum batteries.
\newblock {\em Physical Review A}, 111(1):012204, 2025.

\bibitem{Song2025SelfDischargeQB}
Wan-Lu Song, Ji-Ling Wang, Bin Zhou, Wan-Li Yang, and Jun-Hong An.
\newblock Self-discharging mitigated quantum battery.
\newblock {\em Physical Review Letters}, 135(2):020405, 2025.

\bibitem{Khoudiri2025CoherenceDrivenQATM}
Achraf Khoudiri, Abderrahman Oularabi, Khadija El~Anouz, {\.I}lkay Demir, and Abderrahim El~Allati.
\newblock Coherence-driven quantum battery charging via autonomous thermal machines: Energy transfer, memory effects, and ergotropy enhancement.
\newblock {\em arXiv preprint}, 2025.

\bibitem{2023JPhA...56A5302K}
F.~H. {Kamin}, Z.~{Abuali}, H.~{Ness}, and S.~{Salimi}.
\newblock {Quantum battery charging by non-equilibrium steady-state currents}.
\newblock {\em Journal of Physics A Mathematical General}, 56(27):275302, July 2023.

\bibitem{Centrone2023}
Federico Centrone, Luca Mancino, and Mauro Paternostro.
\newblock Charging batteries with quantum squeezing.
\newblock {\em Phys. Rev. A}, 108:052213, Nov 2023.

\bibitem{QUTE2024_CavityScheme}
S.~Yumul et~al.
\newblock Proposed scheme for a cavity-based quantum battery.
\newblock {\em Adv. Quantum Technol.}, 2024.
\newblock Early View.

\bibitem{rzgc-h78v}
Jin-Tian Zhang, Cheng-Ge Liu, and Qing Ai.
\newblock Improving quantum-battery lifetime by the electromagnetically-induced-transparency effect and bound states.
\newblock {\em Phys. Rev. A}, 112:043705, Oct 2025.

\bibitem{Hadipour2025BathModulationQB}
Maryam Hadipour and Soroush Haseli.
\newblock Nonequilibrium quantum batteries: Amplified work extraction through thermal bath modulation.
\newblock {\em arXiv preprint arXiv:2502.05508}, 2025.

\bibitem{Utsumi2010}
Y.~Utsumi, D.~Golubev, M.~Marthaler, K.~Saito, T.~Fujisawa, and G.~Sch\"on.
\newblock Bidirectional single-electron counting and the fluctuation theorem.
\newblock {\em Phys. Rev. B}, 81:125331, 2010.

\bibitem{Hernandez2021}
S.~Hern\'andez-G\'omez, N.~Staudenmaier, M.~Campisi, and N.~Fabbri.
\newblock Cumulant generating function for work, heat and energy in driven open quantum systems.
\newblock {\em New J. Phys.}, 23(6):065004, 2021.

\bibitem{Esposito2009}
M.~Esposito, U.~Harbola, and S.~Mukamel.
\newblock Nonequilibrium fluctuations, fluctuation theorems, and counting statistics in quantum systems.
\newblock {\em Rev. Mod. Phys.}, 81:1665--1702, 2009.

\bibitem{Touchette2009}
H.~Touchette.
\newblock The large deviation approach to statistical mechanics.
\newblock {\em Phys. Rep.}, 478:1--69, 2009.

\bibitem{Paulino2024PRL_AdiabaticFCS}
P.~J. Paulino et~al.
\newblock Large deviation full counting statistics in adiabatic open quantum dynamics.
\newblock {\em Phys. Rev. Lett.}, 132:260402, 2024.

\bibitem{Joshi2025PRL_FCS_Ions}
L.~K. Joshi et~al.
\newblock Measuring full counting statistics in a trapped-ion quantum simulator.
\newblock {\em Phys. Rev. Lett.}, 2025.
\newblock in press.

\bibitem{Rodriguez2023}
Carla Rodriguez, Dario Rosa, and Jan Olle.
\newblock Artificial intelligence discovery of a charging protocol in a micromaser quantum battery.
\newblock {\em Phys. Rev. A}, 108(4):042618, 2023.

\bibitem{Erdman2024}
Paolo~Andrea Erdman, Gian~Marcello Andolina, Vittorio Giovannetti, and Frank No{\'e}.
\newblock Reinforcement learning optimization of the charging of a dicke quantum battery.
\newblock {\em Phys. Rev. Lett.}, 133(24):243602, 2024.

\bibitem{Hoang2024PRR_VQErgo}
D.~T. Hoang, F.~Metz, A.~Thomasen, T.~D. Anh-Tai, T.~Busch, and T.~Fogarty.
\newblock Variational quantum algorithm for ergotropy estimation in quantum many-body batteries.
\newblock {\em Phys. Rev. Research}, 6:013038, 2024.

\bibitem{Sentz2025}
P.~Sentz, S.~Nicholson, Y.~Cho, S.~Reddy, B.~Keith, and S.~G\"unther.
\newblock Learning thermodynamic master equations for open quantum systems.
\newblock {\em arXiv preprint arXiv:2506.01882}, 2025.

\bibitem{Du2025MeasQB_PRR}
J.~Du, X.~Liu, W.~Zhang, Q.~Zhao, and F.-Q. Dou.
\newblock Nonequilibrium quantum battery based on quantum measurements.
\newblock {\em Physical Review Research}, 7:013151, 2025.

\bibitem{Rodriguez2024OQC_NJP}
Ricard~Ravell Rodr{\'\i}guez, Behnam Ahmadi, Gabriel Su{\'a}rez, Pawe{\l} Mazurek, Shabir Barzanjeh, Pawe{\l} Horodecki, et~al.
\newblock Optimal quantum control of charging quantum batteries.
\newblock {\em New Journal of Physics}, 26(4):043004, 2024.

\bibitem{Santos2021SelfDischargePRE}
Alan~C. Santos.
\newblock Quantum advantage of two-level batteries in the self-discharging process.
\newblock {\em Physical Review E}, 103:042118, 2021.

\bibitem{PhysRevA.86.043843}
Saar Rahav, Upendra Harbola, and Shaul Mukamel.
\newblock Heat fluctuations and coherences in a quantum heat engine.
\newblock {\em Phys. Rev. A}, 86:043843, Oct 2012.

\bibitem{PhysRevA.110.052214}
Manash~Jyoti Sarmah and Himangshu~Prabal Goswami.
\newblock Efficiency fluctuations of a heat engine with noise-induced quantum coherences.
\newblock {\em Phys. Rev. A}, 110:052214, Nov 2024.

\bibitem{scully2011quantum}
Marlan~O Scully, Kimberly~R Chapin, Konstantin~E Dorfman, Moochan~Barnabas Kim, and Anatoly Svidzinsky.
\newblock Quantum heat engine power can be increased by noise-induced coherence.
\newblock {\em Proceedings of the National Academy of Sciences}, 108(37):15097--15100, 2011.

\bibitem{PhysRevA.88.013842}
Himangshu~Prabal Goswami and Upendra Harbola.
\newblock Thermodynamics of quantum heat engines.
\newblock {\em Phys. Rev. A}, 88:013842, Jul 2013.

\bibitem{PhysRevResearch.4.L032034}
Jaegon Um, Konstantin~E. Dorfman, and Hyunggyu Park.
\newblock Coherence-enhanced quantum-dot heat engine.
\newblock {\em Phys. Rev. Res.}, 4:L032034, Aug 2022.

\bibitem{svidzinsky2012enhancing}
Anatoly~A Svidzinsky, Konstantin~E Dorfman, and Marlan~O Scully.
\newblock Enhancing photocell power by noise-induced coherence.
\newblock {\em Coherent Optical Phenomena}, 1:7--24, 2012.

\bibitem{PhysRevA.107.052217}
Manash~Jyoti Sarmah and Himangshu~Prabal Goswami.
\newblock Work flux and efficiency at maximum power of a triply squeezed engine.
\newblock {\em Phys. Rev. A}, 107:052217, May 2023.

\bibitem{PhysRevLett.119.050602}
Yueyang Zou, Yue Jiang, Yefeng Mei, Xianxin Guo, and Shengwang Du.
\newblock Quantum heat engine using electromagnetically induced transparency.
\newblock {\em Phys. Rev. Lett.}, 119:050602, Aug 2017.

\bibitem{sarmah2024noise}
Manash~Jyoti Sarmah and Himangshu~Prabal Goswami.
\newblock Noise-induced coherent ergotropy of a quantum heat engine.
\newblock {\em Physical Review A}, 110(3):032213, 2024.

\bibitem{SARMAH2023129135}
Manash~Jyoti Sarmah and Himangshu~Prabal Goswami.
\newblock Learning coherences from nonequilibrium fluctuations in a quantum heat engine.
\newblock {\em Physica A: Statistical Mechanics and its Applications}, 627:129135, 2023.

\bibitem{Varoquaux2017CV}
G.~Varoquaux, P.~R. Raamana, D.~A. Engemann, A.~Hoyos-Idrobo, Y.~Schwartz, and B.~Thirion.
\newblock Assessing and tuning brain decoders: cross-validation, caveats, and guidelines.
\newblock {\em NeuroImage}, 2017.

\bibitem{Kaufman2012Leakage}
S.~Kaufman, S.~Rosset, C.~Perlich, and O.~Stitelman.
\newblock Leakage in data mining: Formulation, detection, and avoidance.
\newblock {\em ACM Transactions on Knowledge Discovery from Data}, 2012.

\bibitem{Rumelhart1986Backprop}
David~E. Rumelhart, Geoffrey~E. Hinton, and Ronald~J. Williams.
\newblock Learning representations by back-propagating errors.
\newblock {\em Nature}, 323:533--536, 1986.

\bibitem{Friedman2001GBM}
Jerome~H. Friedman.
\newblock Greedy function approximation: A gradient boosting machine.
\newblock {\em The Annals of Statistics}, 29(5):1189--1232, 2001.

\bibitem{Breiman2001RandomForests}
Leo Breiman.
\newblock Random forests.
\newblock {\em Machine Learning}, 45(1):5--32, 2001.

\bibitem{gorishniy2021revisiting}
Yury Gorishniy, Ivan Rubachev, Valentin Khrulkov, and Artem Babenko.
\newblock Revisiting deep learning models for tabular data.
\newblock {\em Advances in neural information processing systems}, 34:18932--18943, 2021.

\bibitem{hollmann2025accurate}
Noah Hollmann, Samuel M{\"u}ller, Lennart Purucker, Arjun Krishnakumar, Max K{\"o}rfer, Shi~Bin Hoo, Robin~Tibor Schirrmeister, and Frank Hutter.
\newblock Accurate predictions on small data with a tabular foundation model.
\newblock {\em Nature}, 637(8045):319--326, 2025.

\end{thebibliography}

\end{document}